\documentclass{article}
\usepackage{graphicx}  % standard LaTeX graphics tool;
\usepackage{amsmath}   % For getting proper math eqns
\usepackage[compress]{cite}
\usepackage{amssymb}   % Used for getting various symbols eg. gtrsim, lesssim etc.
\usepackage{bm} % For bold math (esp of lower greek letters)
\usepackage{dcolumn}% Align table columns on decimal point
\usepackage{color}
\usepackage{mathrsfs}
\usepackage{amsfonts}
\RequirePackage[colorlinks,citecolor=blue,urlcolor=magenta,linkcolor=blue]{hyperref}
\def\eq#1{{Eq.~(\ref{#1})}}
\def\eqs#1{{Eqs.~(\ref{#1})}}
\def\sect#1{{Sec.~\ref{#1}}}
\title{Metric factorizability and equivalence of brane world models with Brans-Dicke theory}
\author{Sumanta Chakraborty 
\footnote{sumantac.physics@gmail.com}
\footnote{sumanta@iucaa.ernet.in}\\
{\small {IUCAA, Post Bag 4, Ganeshkhind, Pune University Campus, Pune 411007, India}}\\
and\\
Soumitra SenGupta
\footnote{tpssg@iacs.res.in}\\
{\small{Department of Theoretical Physics, Indian Association for the Cultivation of Science, Kolkata 700032, India}}}
\date{\today}  %% This command  will supress printing the date. 
\begin{document}

\maketitle

%%%%%%%%%%%%%%%%%%%%%%%%%%%%%%%%%%%%%%%%%%%%%%%%%%%%%%%%%%%%%%%%%%%%%%%%%%%%%%%%%%%%%%%%%%%%%%%%%
%%%%%%%%%%%%%%%%%%%%%%%%%%%%%%%%%%%%%%%%%%%%%%%%%%%%%%%%%%%%%%%%%%%%%%%%%%%%%%%%%%%%%%%%%%%%%%%%%
%%%%%%%%%%%%%%%%%%%%%%%%%%%%%%%%%%%%%%%%%%%%%%%%%%%%%%%%%%%%%%%%%%%%%%%%%%%%%%%%%%%%%%%%%%%%%%%%%
\begin{abstract}

In the standard brane world models, the bulk metric ansatz is usually assumed to be factorizable in brane and bulk coordinates. However, it is \emph{not} self-evident that it is always possible to factorize the bulk metric. Using the gradient expansion scheme, which involves the expansion of bulk quantities in terms of the brane-to-bulk curvature ratio, as a perturbative parameter, we explicitly show that metric factorizability is a \emph{valid} assumption upto second order in the perturbative expansion. We also argue from our result that the same should be true for \emph{all} orders in the perturbation scheme. We further establish that the nonlocal terms present in the bulk gravitational field equation can be replaced by the radion field; the effective action on the brane thereby obtained resembles the Brans-Dicke theory of gravity. 

\end{abstract}
%%%%%%%%%%%%%%%%%%%%%%%%%%%%%%%%%%%%%%%%%%%%%%%%%%%%%%%%%%%%%%%%%%%%%%%%%%%%%%%%%%%%%%%%%%%%%%%%
%%%%%%%%%%%%%%%%%%%%%%%%%%%%%%%%%%%%%%%%%%%%%%%%%%%%%%%%%%%%%%%%%%%%%%%%%%%%%%%%%%%%%%%%%%%%%%%%
%%%%%%%%%%%%%%%%%%%%%%%%%%%%%%%%%%%%%%%%%%%%%%%%%%%%%%%%%%%%%%%%%%%%%%%%%%%%%%%%%%%%%%%%%%%%%%%%%
%\newpage
%%%%%%%%%%%%%%%%%%%%%%%%%%%%%%%%%%%%%%%%%%%%%%%%%%%%%%%%%%%%%%%%%%%%%%%%%%%%%%%%%%%%%%%%%%%%%%%%
%%%%%%%%%%%%%%%%%%%%%%%%%%%%%%%%%%%%%%%%%%%%%%%%%%%%%%%%%%%%%%%%%%%%%%%%%%%%%%%%%%%%%%%%%%%%%%%%
%\tableofcontents
%\newpage
%%%%%%%%%%%%%%%%%%%%%%%%%%%%%%%%%%%%%%%%%%%%%%%%%%%%%%%%%%%%%%%%%%%%%%%%%%%%%%%%%%%%%%%%%%%%%%%%%%
%%%%%%%%%%%%%%%%%%%%%%%%%%%%%%%%%%%%%%%%%%%%%%%%%%%%%%%%%%%%%%%%%%%%%%%%%%%%%%%%%%%%%%%%%%%%%%%%%%
%%%%%%%%%%%%%%%%%%%%%%%%%%%%%%%%%%%%%%%%%%%%%%%%%%%%%%%%%%%%%%%%%%%%%%%%%%%%%%%%%%%%%%%%%%%%%%%%%%
\section{Introduction}

The conjecture of the existence of more than four spacetime dimensions has serious implications in high-energy physics. Such higher-dimensional spacetimes appear quite naturally in the context of string theory. There has recently been progress in this regime, especially for theories with extra spatial dimensions. The common perception for all these theories corresponds to the fact that gravity can access the whole of spacetime including the extra dimensions (together known as the bulk), while the standard model fields are localized on four-dimensional submanifold (known as the brane). One of the main motivations behind these models, has been to explain the large hierarchy between the Planck scale ($M_{P}\sim 10^{18}\textrm{GeV}$) and the electroweak scale ($m_{W}\sim 10^{2}\textrm{GeV}$).

First such model was proposed by Arkani-Hamed \textit{et al.} \cite{Hamed1998a,Hamed1998b}. In this model the extra dimensions were assumed to be large, such that the five-dimensional Planck scale differs from the four-dimensional Planck scale by a factor of the volume of these extra dimensions. Thus by assuming more than one extra dimension and a large volume (though still within experimental bounds), the five-dimensional Planck scale can be brought down to the four-dimensional electroweak scale. However, in this case the extra dimensions are assumed to be flat.

From the gravitational viewpoint it is more tempting to take the bulk geometry as warped, with the brane(s) as flat. This was first realized in a setup proposed by Randall and Sundrum (RS)\cite{Randall1999a}, where two branes were held fixed at orbifold fixed points with $S^{1}/Z_{2}$ symmetry. Because of exponential warping the Planck scale in one brane (the Planck brane) was brought down to the electroweak scale in the other brane, known as the visible brane. Such a warped model was also extended to one brane with an infinitely extended bulk \cite{Randall1999b}. In this work, however, we focus on the two branes warped geometry model.

The separation between the branes in the RS model may not be constant and needs to be stabilized. Such a stabilization mechanism was proposed in \cite{Wise1999a,Wise1999b}, while the stabilization for a time-dependent scenario was discussed in \cite{Chakraborty2014a}. The particle phenomenology of various matter fields in this scenario was discussed in \cite{Rizzo2000a,Rizzo2001,Rizzo2000b,SenGupta2009,SenGupta2011,Chakraborty2014b}, with interesting consequences. Recently, these ideas have also been put forward in the context of various alternative gravity theories \cite{Borzou2009,Sepangi2005,Afonso2007,Chakraborty2014c,Chakraborty2014d,Chakraborty2015a}.

All these results depend on a crucial fact, the factorizability of metric ansatz. However. there are objections against this assumption of factorizability; further, it is also not self evident, why the metric ansatz should be factorizable \cite{Alwis2004}. In this work, we have tried to address this issue using low-energy effective action obtained by solving the bulk equations. The bulk equations in general are not exactly solvable; a convenient way to handle the situation at low energy is to expand the bulk variables in terms of the ratio of four-dimensional curvature to bulk curvature. This method, known as the gradient expansion method was developed by Kanno and Soda \cite{Kanno2002a,Kanno2002b,Kanno2003,Kanno2004}. In \cite{Kanno2005} the gradient expansion method has been used up to first order to show that the factorizable metric ansatz is valid up to linear in this perturbative expansion. In this work we obtain the second-order correction to the metric in this gradient expansion scheme, which leads to the effective action up to second order. This also exhibits the factorizable nature, which in turn enables us to generalize our result to include higher-order corrections. We conclude that at any order the metric \emph{is} factorizable; thus. factorizability of the metric is a valid assumption.

Along with the issue of factorizability of the metric ansatz, we also address the equivalence of this bulk-brane system with the scalar-tensor or Brans-Dicke theory of gravity. The solutions to bulk equations intrinsically inherit nonlocal terms which, as we have argued, can be traded off through the radion field. This equivalence was shown earlier in \cite{Kanno2003} for first-order perturbative corrections through the gradient expansion method. We have reformulated the previous method and show explicitly that up to second order of perturbative expansion, when the nonlocal terms are eliminated, the field equation on the brane becomes local and equivalent to that of the Brans-Dicke theory of gravity. We also argue that this result can be generalized to arbitrary higher orders in the perturbative expansion. The same assertion also follows from the effective action; i.e., the effective action can be written explicitly in the Brans-Dicke form. 

The paper is organized as follows: In \sect{Fact.Sec.02} we review the gradient expansion method and evaluate the second-order correction to the bulk metric. Then, in \sect{Fact.Sec.03} we use the bulk metric in order to determine the effective action and the equation of motion it corresponds to. Along with these, we also present the criteria for obtaining the second-order field equation from this effective action. Finally, in \sect{Fact.Sec.04} we establish the equivalence of this bulk-brane system with the Brans-Dicke theory of gravity. We then conclude with a short discussion of our results. 

In this work we will set $c$ and $\hbar$ to unity. The Latin indices $a,b,\ldots$ run over the full spacetime, while Greek indices $\mu ,\nu ,\ldots$ represent the brane coordinates. The metric signature is taken to be $(-,+,+,+,+,+)$.

%%%%%%%%%%%%%%%%%%%%%%%%%%%%%%%%%%%%%%%%%%%%%%%%%%%%%%%%%%%%%%%%%%%%%%%%%%%%%%%%%%%%%%%%%%%%%%%%%%%%
%%%%%%%%%%%%%%%%%%%%%%%%%%%%%%%%%%%%%%%%%%%%%%%%%%%%%%%%%%%%%%%%%%%%%%%%%%%%%%%%%%%%%%%%%%%%%%%%%%%%
%%%%%%%%%%%%%%%%%%%%%%%%%%%%%%%%%%%%%%%%%%%%%%%%%%%%%%%%%%%%%%%%%%%%%%%%%%%%%%%%%%%%%%%%%%%%%%%%%%%%
\section{Gradient Expansion and Higher Order Terms}\label{Fact.Sec.02}

The metric ansatz for the five-dimensional spacetime is taken in Gaussian normal coordinates, where we denote the brane coordinates by $x^{\mu}$ and the bulk coordinate by $y$ such that 
%%%%%%%%%%%%%%%%%%%%%%%%%%%%%%%%%%%%%%%%%%%%%%%%%%%%%
\begin{equation}\label{Fact.Sec.02.01}
ds^{2}=h_{\mu \nu}(y,x^{\mu})dx^{\mu}dx^{\nu}+dy^{2}.
\end{equation}
%%%%%%%%%%%%%%%%%%%%%%%%%%%%%%%%%%%%%%%%%%%%%%%%%%%%%%%
Thus, the metric in general is not taken as factorizable. The branes are assumed to be moving in the coordinate chart where they are placed at
%%%%%%%%%%%%%%%%%%%%%%%%%%%%%%%%%%%%%%%%%%%%%%%%%%%%%%%%
\begin{equation}\label{Fact.Sec.02.02}
y=\phi _{+}(x^{\mu});\qquad y=\phi _{-}(x^{\mu}),
\end{equation}
%%%%%%%%%%%%%%%%%%%%%%%%%%%%%%%%%%%%%%%%%%%%%%%%%%%%%%%%%%
and in the literature they are often quoted as moduli fields. In order to determine the brane geometry we need to solve the bulk equations. The form of the metric ansatz suggests that the extrinsic curvature on $y=\textrm{constant}$ hypersurface can be found through its decomposition into traceless and trace part as
%%%%%%%%%%%%%%%%%%%%%%%%%%%%%%%%%%%%%%%%%%%%%%%%%%%%%%%%%%%
\begin{equation}\label{Fact.Sec.02.03}
K_{\mu \nu}=-\frac{1}{2}\dfrac{\partial h_{\mu \nu}}{\partial y},\qquad
K_{\mu \nu}=\Sigma _{\mu \nu}+\frac{1}{4}h_{\mu \nu}K,\qquad
K=-\dfrac{\partial \ln \sqrt{-h}}{\partial y}.
\end{equation} 
%%%%%%%%%%%%%%%%%%%%%%%%%%%%%%%%%%%%%%%%%%%%%%%%%%%%%%%%%%%%%%%%
Using these properties of extrinsic curvature in the bulk equations lead to the equations \cite{Kanno2002a,Kanno2002b,Kanno2003,Kanno2004}
%%%%%%%%%%%%%%%%%%%%%%%%%%%%%%%%%%%%%%%%%%%%%%%%%%%%%%%%%%%%%%%%%
\begin{align}
\partial _{y}\Sigma _{\mu \nu}-K\Sigma _{\mu \nu}&=-\left[R_{\mu \nu}(h)-\frac{1}{4}h_{\mu \nu}R(h)\right]
\label{Fact.Sec.02.04}
\\
\frac{3}{4}K^{2}-\Sigma ^{\alpha \beta}\Sigma _{\alpha \beta}&=R(h)+\frac{12}{\ell ^{2}}
\label{Fact.Sec.02.05}
\\
\nabla _{\nu}\Sigma _{\mu}^{\nu}&-\frac{3}{4}\nabla _{\mu}K=0,
\label{Fact.Sec.02.06}
\end{align}
%%%%%%%%%%%%%%%%%%%%%%%%%%%%%%%%%%%%%%%%%%%%%%%%%%%%%%%%%%%%%%%%%%%
where the covariant derivatives are with respect to the metric $h_{\mu \nu}$, and all the curvature components, i.e., Ricci tensor and Ricci scalar, are to be determined using $h_{\mu \nu}$. In general we should first solve \eq{Fact.Sec.02.04} and integrate over $y$ to get $\Sigma _{\mu \nu}$, and then we may solve for $K$ from \eq{Fact.Sec.02.05} to get $K_{\mu \nu}$, which finally can be integrated to obtain $h_{\mu \nu}$. However as the curvature components depend on $h_{\mu \nu}$, this procedure cannot in general be implemented. This poses a serious problem; this can be bypassed by observing that we are seeking a low-energy effective theory, where the brane matter energy density can be assumed to be much smaller compared to the bulk cosmological constant. This implies that the four-dimensional curvature is much smaller compared to the five-dimensional one and the gradient expansion scheme can be applicable \cite{Kanno2002a,Kanno2002b,Kanno2003,Kanno2004}. 

At zeroth order, the curvature terms can be neglected in comparison to the extrinsic curvature terms. Being isotropic at this order, the anisotropic term $\Sigma _{\mu \nu}$ vanishes. Then, the metric at zeroth order is $h_{\mu \nu}=a^{2}(y)g_{\mu \nu}(x)$, with the standard warp factor $a(y)=e^{-y/\ell}$. This iteration scheme helps to write the metric $h_{\mu \nu}$ as a sum of tensors constructed from $g_{\mu \nu}$. Thus the metric has the form of a perturbative series expansion,
%%%%%%%%%%%%%%%%%%%%%%%%%%%%%%%%%%%%%%%%%%%%%%%%%%%%%%%%%%%%%%%%%%%%%%%%%%%%%5
\begin{align}\label{Fact.Sec.02.07}
h_{\mu \nu}=a^{2}(y)\left[g_{\mu \nu}(x)+f_{\mu \nu}(y,x)+q_{\mu \nu}(y,x)+\cdots \right];\qquad 
a(y)=e^{-y/\ell}.
\end{align}
%%%%%%%%%%%%%%%%%%%%%%%%%%%%%%%%%%%%%%%%%%%%%%%%%%%%%%%%%%%%%%%%%%%%%%%%%%%%%%%
where $f_{\mu \nu}(y,x)$ corresponds to leading-order correction, and $q_{\mu \nu}$ represents the second-order correction. After calculating second-order corrections a pattern will emerge from which the effective action can be determined at all orders. We will elaborate on this at a later stage. 

In a similar manner, we can expand both the extrinsic curvature and the trace-free part as
%%%%%%%%%%%%%%%%%%%%%%%%%%%%%%%%%%%%%%%%%%%%%%%%%%%%%%%%%%%%%%%%%%%%%%%%%%%%%
\begin{align}
K^{\mu}_{\nu}&=\frac{1}{\ell}\delta ^{\mu}_{\nu}+K^{(1)\mu}_{\nu}+K^{(2)\mu}_{\nu}+\cdots
\label{Fact.Sec.02.08}
\\
\Sigma ^{\mu}_{\nu}&=0+\Sigma ^{(1)\mu}_{\nu}+\Sigma ^{(2)\mu}_{\nu}.
\label{Fact.Sec.02.09}
\end{align}
%%%%%%%%%%%%%%%%%%%%%%%%%%%%%%%%%%%%%%%%%%%%%%%%%%%%%%%%%%%%%%%%%%%%%%%%%%%%
In the above expansion, objects with superscript $(1)$ denote first-order corrections, while those with superscript $(2)$ denote second-order corrections, and so on. We briefly discuss the first-order formulation, leading to a possible solution for $f_{\mu \nu}$, then we shall elaborate on the second-order calculation in order to obtain the tensor $q_{\mu \nu}$. These will be used later to get the effective action. 

%%%%%%%%%%%%%%%%%%%%%%%%%%%%%%%%%%%%%%%%%%%%%%%%%%%%%%%%%%%%%%%%%%%%%%%%%%%%%%%
%%%%%%%%%%%%%%%%%%%%%%%%%%%%%%%%%%%%%%%%%%%%%%%%%%%%%%%%%%%%%%%%%%%%%%%%%%%%%%%
%%%%%%%%%%%%%%%%%%%%%%%%%%%%%%%%%%%%%%%%%%%%%%%%%%%%%%%%%%%%%%%%%%%%%%%%%%%%%%%
\subsection{First Order}\label{Grad.Sec.02.First}

The first-order equations are obtained by considering terms in which $K^{(1)\mu}_{\nu}$ and $\Sigma ^{(1)\mu}_{\nu}$ appear once in the expressions. For example, $K^{2}=(16/\ell ^{2})+(8/\ell)K^{(1)}$, where we have used the result that at zeroth order $K^{(0)}=(4/\ell)$. Similar considerations apply to $\Sigma ^{\mu}_{\nu}$ as well, with the fact that at zeroth order it vanishes. Thus, the bulk equations at first order take the forms \cite{Kanno2002a,Kanno2002b,Kanno2003,Kanno2004}
%%%%%%%%%%%%%%%%%%%%%%%%%%%%%%%%%%%%%%%%%%%%%%%%%%%%%%%%%%%%%%%%%%%%%%%
\begin{align}
\partial _{y}\Sigma ^{(1)\mu}_{\nu}-(4/\ell)\Sigma ^{(1)\mu}_{\nu}&=-\left[R^{(1)\mu}_{\nu}(h)-\frac{1}{4}\delta ^{\mu}_{\nu}R^{(1)}(h)\right]
\label{Fact.Sec.02.10}
\\
\frac{6}{\ell}K^{(1)}&=R^{(1)}(h)
\label{Fact.Sec.02.11}
\\
\nabla _{\nu}\Sigma _{(1)\mu}^{\nu}&-\frac{3}{4}\nabla _{\mu}K^{(1)}=0.
\label{Fact.Sec.02.12}
\end{align}
%%%%%%%%%%%%%%%%%%%%%%%%%%%%%%%%%%%%%%%%%%%%%%%%%%%%%%%%%%%%%%%%%%%%%%%%%
Here, the covariant derivatives are with respect to the metric $g_{\mu \nu}$, and $R^{(1)}(h)$ imply the Ricci scalar calculated using $a^{2}(y)g_{\mu \nu}$. Similar conclusions can be reached for the Ricci tensor as well. For this reason, we will henceforth provide the curvature components with respect to the metric $g_{\mu \nu}$ only, with $a^{2}(y)$ taken out. This reduces the first-order equation (\ref{Fact.Sec.02.11}) to the form
%%%%%%%%%%%%%%%%%%%%%%%%%%%%%%%%%%%%%%%%%%%%%%%%%%%%%%%%%%%%%%%%%%%%%%%%%%
\begin{equation}\label{Fact.Sec.02.13}
 K^{(1)}=\frac{\ell}{6a^{2}}R(g).
\end{equation}
%%%%%%%%%%%%%%%%%%%%%%%%%%%%%%%%%%%%%%%%%%%%%%%%%%%%%%%%%%%%%%%%%%%%%%%%%%%%%
Similarly, integrating over $y$ in \eq{Fact.Sec.02.10} leads to the first-order traceless part of the extrinsic curvature as
%%%%%%%%%%%%%%%%%%%%%%%%%%%%%%%%%%%%%%%%%%%%%%%%%%%%%%%%%%%%%%%%%%%%%%%%%%%%
\begin{align}
\Sigma ^{(1)\mu}_{\nu}&=\frac{\ell}{2a^{2}}\left(R^{\mu}_{\nu}(g)-\frac{1}{4}\delta ^{\mu}_{\nu}R(g)\right)
+\frac{1}{a^{4}}\chi ^{\mu}_{\nu}(x)
\label{Fact.Sec.02.14}
\\
\chi ^{\mu}_{\mu}&=0,\qquad \nabla _{\mu}\chi ^{\mu}_{\nu}=0,
\label{Fact.Sec.02.15}
\end{align}
%%%%%%%%%%%%%%%%%%%%%%%%%%%%%%%%%%%%%%%%%%%%%%%%%%%%%%%%%%%%%%%%%%%%%%%%%%%%%
where in \eq{Fact.Sec.02.14} $\chi _{\mu}^{\nu}$ is an arbitrary constant of integration, which, due to the traceless property of $\Sigma ^{\mu}_{\nu}$ and \eq{Fact.Sec.02.12}, satisfies the last two relations in \eq{Fact.Sec.02.15}. From now on we will drop the argument of curvature components for notational convenience; every curvature component will be assumed to be constructed from $g_{\mu \nu}$. Then, from $\Sigma ^{(1)\mu}_{\nu}$ given in \eq{Fact.Sec.02.14} and $K^{(1)}$ provided in \eq{Fact.Sec.02.13}, we can construct $K^{(1)\mu}_{\nu}$; this, after integration over $y$ coordinate, leads to the corrected metric up to first order as
%%%%%%%%%%%%%%%%%%%%%%%%%%%%%%%%%%%%%%%%%%%%%%%%%%%%%%%%%%%%%%%%%%%%%%%%%%%%%%%
\begin{equation}\label{Fact.Sec.02.16}
f_{\mu \nu}(y,x)=-\frac{\ell ^{2}}{2a^{2}}\left(R_{\mu \nu}-\frac{1}{6}g_{\mu \nu}R\right)-\frac{\ell}{2a^{4}}\chi _{\mu \nu}(x)+C_{\mu \nu}(x).
\end{equation}
%%%%%%%%%%%%%%%%%%%%%%%%%%%%%%%%%%%%%%%%%%%%%%%%%%%%%%%%%%%%%%%%%%%%%%%%%%%%%%%
Here $C_{\mu \nu}$ is a constant of integration. Using this, the first-order corrected metric $h_{\mu \nu}$ turns out to have the following expression:
%%%%%%%%%%%%%%%%%%%%%%%%%%%%%%%%%%%%%%%%%%%%%%%%%%%%%%%%%%%%%%%%%%%%%%%%%%%%%%%%
\begin{align}\label{Fact.Sec.02.19}
h_{\mu \nu}=a^{2}(y)\left[g_{\mu \nu}-\frac{\ell ^{2}}{2a^{2}}\left(R_{\mu \nu}-\frac{1}{6}g_{\mu \nu}R\right)-\frac{\ell}{2a^{4}}\chi _{\mu \nu}(x)+C_{\mu \nu}(x)\right].
\end{align}
%%%%%%%%%%%%%%%%%%%%%%%%%%%%%%%%%%%%%%%%%%%%%%%%%%%%%%%%%%%%%%%%%%%%%%%%%%%%%%%%%5
As an aside we would like to point out a particular situation in which case one of the arbitrary constants can be obtained uniquely and our result reduces to that derived in \cite{Kanno2002b}. This condition amounts to fixing the brane positions. Thus, if we assume that the branes are fixed at $y=0$ and $y=\pi$, respectively, and impose the boundary condition that $h_{\mu \nu}(y=0,x)=g_{\mu \nu}$, then we have
%%%%%%%%%%%%%%%%%%%%%%%%%%%%%%%%%%%%%%%%%%%%%%%%%%%%%%%%%%%%%%%%%%%%%%%%%%%%%%%%%5
\begin{align}\label{Fact.Sec.02.17}
C_{\mu \nu}(x)=(\ell ^{2}/2)\left[R_{\mu \nu}-(1/6)g_{\mu \nu}R\right]+(\ell/2)\chi _{\mu \nu}(x).
\end{align}
%%%%%%%%%%%%%%%%%%%%%%%%%%%%%%%%%%%%%%%%%%%%%%%%%%%%%%%%%%%%%%%%%%%%%%%%%%%
Thus, in this particular situation with the above boundary condition, we obtain the first-order correction as
%%%%%%%%%%%%%%%%%%%%%%%%%%%%%%%%%%%%%%%%%%%%%%%%%%%%%%%%%%%%%%%%%%%%%%%%%%%%%%%%%
\begin{equation}\label{Fact.Sec.02.18}
f_{\mu \nu}(y,x)=\frac{\ell ^{2}}{2}\left(1-\frac{1}{a^{2}}\right)\left(R_{\mu \nu}-\frac{1}{6}g_{\mu \nu}R\right)+\frac{\ell}{2}\left(1-\frac{1}{a^{4}}\right)\chi _{\mu \nu}(x).
\end{equation}
%%%%%%%%%%%%%%%%%%%%%%%%%%%%%%%%%%%%%%%%%%%%%%%%%%%%%%%%%%%%%%%%%%%%%%%%%%%%%%%%
Note that this matches exactly with the one obtained in \cite{Kanno2002b}. However, in this work we want to keep the brane positions variable; thus we will work with \eq{Fact.Sec.02.19}, which differs from the choice in \cite{Kanno2002b}.
Having obtained the metric with the first-order correction term included, we now proceed to calculate the second-order correction in greater detail.

%%%%%%%%%%%%%%%%%%%%%%%%%%%%%%%%%%%%%%%%%%%%%%%%%%%%%%%%%%%%%%%%%%%%%%%%%%%%%%%%%
%%%%%%%%%%%%%%%%%%%%%%%%%%%%%%%%%%%%%%%%%%%%%%%%%%%%%%%%%%%%%%%%%%%%%%%%%%%%%%%%%
%%%%%%%%%%%%%%%%%%%%%%%%%%%%%%%%%%%%%%%%%%%%%%%%%%%%%%%%%%%%%%%%%%%%%%%%%%%%%%%5%
\subsection{Second Order}

At second order the bulk equations contain a single power of second-order objects, double power of first-order objects, and so on. For example, at second order our expression would include only $\Sigma ^{(1)\mu}_{\nu}$, but we can have terms like $K^{(1)\mu}_{\nu}\Sigma ^{(1)\mu}_{\nu}$. Thus at second order the bulk equations \eqs{Fact.Sec.02.04}-(\ref{Fact.Sec.02.06}) reduce to the following forms \cite{Kanno2002a,Kanno2002b,Kanno2003,Kanno2004}:
%%%%%%%%%%%%%%%%%%%%%%%%%%%%%%%%%%%%%%%%%%%%%%%%%%%%%%%%%%%%%%%%%%%%%%%%%%%%%%%%
\begin{align}
\partial _{y}\Sigma ^{(2)\mu}_{\nu}&-\frac{4}{\ell}\Sigma ^{(2)\mu}_{\nu}=
-\left(R^{(2)\mu}_{\nu}-\frac{1}{4}R^{(2)}\delta ^{\mu}_{\nu}\right)+K^{(1)}\Sigma ^{(1)\mu}_{\nu}
\label{Fact.Sec.02.20}
\\
K^{(2)}&=\frac{\ell}{6}\left[-\frac{3}{4}\left(K^{(1)}\right)^{2}
+\Sigma ^{(1)\alpha}_{\beta}\Sigma ^{(1)\beta}_{\alpha}+R^{(2)}\right]
\label{Fact.Sec.02.21}
\\
\partial _{y}K^{(2)}&-\frac{2}{\ell}K^{(2)}=\frac{1}{4}\left(K^{(1)}\right)^{2}
+\Sigma ^{(1)\mu}_{\nu}\Sigma ^{(1)\nu}_{\mu}.
\label{Fact.Sec.02.22}
\end{align}
%%%%%%%%%%%%%%%%%%%%%%%%%%%%%%%%%%%%%%%%%%%%%%%%%%%%%%%%%%%%%%%%%%%%%%%%%%%%%%%%%%%%%%%%5
In order to obtain the Ricci tensor and scalar at second order we should use the metric corrected up to the first order, i.e., the result provided in \eq{Fact.Sec.02.19}. Thus, in the second order we have the following expression:
%%%%%%%%%%%%%%%%%%%%%%%%%%%%%%%%%%%%%%%%%%%%%%%%%%%%%%%%%%%%%%%%%%%%%%%%
\begin{align}\label{Fact.Sec.02.23}
R^{(2)\alpha}_{\beta}-\frac{1}{4}\delta ^{\alpha}_{\beta}R^{(2)}
&=\frac{\ell ^{2}}{2a^{4}}\Big[R^{\alpha}_{\mu}R^{\mu}_{\beta}-\frac{1}{6}RR^{\alpha}_{\beta}
-\frac{1}{4}\delta ^{\alpha}_{\beta}\left(R^{\mu}_{\nu}R^{\nu}_{\mu}-\frac{1}{6}R^{2}\right)
\nonumber
\\
&-\frac{1}{2}\left(\nabla _{\mu}\nabla _{\beta}R^{\mu \alpha}
+\nabla _{\mu}\nabla ^{\alpha}R^{\mu}_{\beta}\right)
+\frac{1}{3}\nabla ^{\alpha}\nabla _{\beta}R+\frac{1}{2}\square R^{\alpha}_{\beta}
-\frac{1}{12}\delta ^{\alpha}_{\beta}\square R\Big]
\nonumber
\\
&-\frac{\ell}{2a^{6}}\Big[\frac{1}{2}\nabla _{\mu}\nabla _{\beta}\chi ^{\mu \alpha }+\frac{1}{2}\nabla _{\mu}\nabla ^{\alpha}\chi ^{\mu}_{\beta}-\frac{1}{2}\square \chi ^{\alpha}_{\beta}\Big]+\frac{1}{4a^{8}}\Big[\chi ^{\alpha}_{\mu}\chi ^{\mu}_{\beta}-\frac{1}{4}\delta ^{\alpha}_{\beta}\chi ^{\mu \nu}\chi _{\mu \nu}\Big]
\nonumber
\\
&+\frac{1}{a^{2}}\Big[\frac{1}{2}\nabla _{\mu}\nabla _{\beta}C^{\mu \alpha }+\frac{1}{2}\nabla _{\mu}\nabla ^{\alpha}C^{\mu}_{\beta}-\frac{1}{2}\nabla ^{\alpha}\nabla _{\beta}C^{\mu}_{\mu}-\frac{1}{2}\square C^{\alpha}_{\beta}+\frac{1}{\ell ^{2}}C^{\alpha}_{\mu}C^{\mu}_{\beta}
\nonumber
\\
&-\frac{1}{4}\delta ^{\alpha}_{\beta}\left(\frac{1}{\ell ^{2}}C^{\mu \nu}C_{\mu \nu}+\nabla _{\mu}\nabla _{\alpha}C^{\mu \alpha}-\square C^{\mu}_{\mu}\right)\Big].
\end{align}
%%%%%%%%%%%%%%%%%%%%%%%%%%%%%%%%%%%%%%%%%%%%%%%%%%%%%%%%%%%%%%%%%%%%%%%%%%
In order to arrive at the above expression we have used the following result that at second order the curvature tensor can be obtained in the local inertial frame and can be written in terms of derivatives of the metric. In the local inertial frame this amounts to $\delta R^{\alpha}_{\beta}=(1/2)[\nabla _{\mu}\nabla _{\beta}\delta g^{\mu \alpha}+\nabla _{\mu}\nabla ^{\alpha}\delta g^{\mu}_{\beta}-\nabla _{\beta}\nabla ^{\alpha}\delta g^{\mu}_{\mu}-\square \delta g^{\alpha}_{\beta}]$. Using $\delta g_{\mu \nu}=f_{\mu \nu}$ from \eq{Fact.Sec.02.19}, we readily obtain most of the terms in the above expression and others come from quadratic combinations. From \eq{Fact.Sec.02.23} we arrive at the expression for Ricci scalar at second order as
%%%%%%%%%%%%%%%%%%%%%%%%%%%%%%%%%%%%%%%%%%%%%%%%%%%%%%%%%%%%%%%%%%%%%%%%%%%
\begin{equation}\label{Fact.Sec.02.24}
R^{(2)}=\frac{\ell ^{2}}{2a^{4}}\left(R^{\mu}_{\nu}R^{\nu}_{\mu}-\frac{1}{6}R^{2}\right)+\frac{1}{4a^{8}}\chi ^{\mu \nu}\chi _{\mu \nu}+\frac{1}{a^{2}}\left(\frac{1}{\ell ^{2}}C_{\mu \nu}C^{\mu \nu}+\nabla _{\mu}\nabla _{\alpha}C^{\mu \alpha}-\square C^{\mu}_{\mu}\right).
\end{equation}
%%%%%%%%%%%%%%%%%%%%%%%%%%%%%%%%%%%%%%%%%%%%%%%%%%%%%%%%%%%%%%%%%%%%
Note that our expression is different from the one obtained in \cite{Kanno2002b}, because in \cite{Kanno2002b} the fixed brane assumption was invoked. As we are interested in the factorizability of the metric ansatz, we have kept the brane positions arbitrary. 

Now using \eq{Fact.Sec.02.21} with the help of the Ricci scalar at second order and $\Sigma ^{\mu}_{\nu}$ at first order, the trace of the extrinsic curvature at second order turns out to be
%%%%%%%%%%%%%%%%%%%%%%%%%%%%%%%%%%%%%%%%%%%%%%%%%%%%%%%%%%%%%%%%%%%%%%
\begin{align}\label{Fact.Sec.02.25}
K^{(2)}&=\frac{\ell}{6}\left[-\frac{3}{4}\left(K^{(1)}\right)^{2}
+\Sigma ^{(1)\alpha}_{\beta}\Sigma ^{(1)\beta}_{\alpha}+R^{(2)}\right]
\nonumber
\\
&=\frac{\ell ^{3}}{8a^{4}}\left(R^{\alpha}_{\beta}R^{\beta}_{\alpha}-\frac{2}{9}R^{2}\right)+\frac{5\ell}{24a^{8}}\chi ^{\mu \nu}\chi _{\mu \nu}+\frac{\ell}{6a^{2}}\left(\frac{1}{\ell ^{2}}C_{\mu \nu}C^{\mu \nu}+\nabla _{\mu}\nabla _{\alpha}C^{\mu \alpha}-\square C^{\mu}_{\mu}\right).
\end{align}
%%%%%%%%%%%%%%%%%%%%%%%%%%%%%%%%%%%%%%%%%%%%%%%%%%%%%%%%%%%%%%%%%%%%%%
The traceless part of the extrinsic curvature can be obtained by integrating \eq{Fact.Sec.02.20} over the extra coordinate, which leads to the following expression:
%%%%%%%%%%%%%%%%%%%%%%%%%%%%%%%%%%%%%%%%%%%%%%%%%%%%%%%%%%%%%%%%%%%%%%%%
\begin{align}\label{Fact.Sec.02.26}
\Sigma ^{(2)\alpha}_{\beta}&=-\frac{\ell ^{2}y}{2a^{4}}S^{\alpha}_{\beta}+\frac{\ell ^{2}}{4a^{6}}\Big[\frac{1}{2}\nabla _{\mu}\nabla _{\beta}\chi ^{\mu \alpha }+\frac{1}{2}\nabla _{\mu}\nabla ^{\alpha}\chi ^{\mu}_{\beta}-\frac{1}{2}\square \chi ^{\alpha}_{\beta}\Big]-\frac{\ell}{16a^{8}}\Big[\chi ^{\alpha}_{\mu}\chi ^{\mu}_{\beta}-\frac{1}{4}\delta ^{\alpha}_{\beta}\chi ^{\mu \nu}\chi _{\mu \nu}\Big].
\nonumber
\\
&+\frac{\ell}{2a^{2}}\Big[\frac{1}{2}\nabla _{\mu}\nabla _{\beta}C^{\mu \alpha }+\frac{1}{2}\nabla _{\mu}\nabla ^{\alpha}C^{\mu}_{\beta}-\frac{1}{2}\nabla ^{\alpha}\nabla _{\beta}C^{\mu}_{\mu}-\frac{1}{2}\square C^{\alpha}_{\beta}+\frac{1}{\ell ^{2}}C^{\alpha}_{\mu}C^{\mu}_{\beta}
\nonumber
\\
&-\frac{1}{4}\delta ^{\alpha}_{\beta}\left(\frac{1}{\ell ^{2}}C^{\mu \nu}C_{\mu \nu}+\nabla _{\mu}\nabla _{\alpha}C^{\mu \alpha}-\square C^{\mu}_{\mu}\right)\Big]+\frac{\ell ^{3}}{a^{4}}t^{\alpha}_{\beta}(x),
\end{align}
%%%%%%%%%%%%%%%%%%%%%%%%%%%%%%%%%%%%%%%%%%%%%%%%%%%%%%%%%%%%%%%%%%%%%%%%%%%
where for convenience we have defined a second-rank tensor $S_{\alpha \beta}$ as \cite{Kanno2002a,Kanno2002b,Kanno2003,Kanno2004}
%%%%%%%%%%%%%%%%%%%%%%%%%%%%%%%%%%%%%%%%%%%%%%%%%%%%%%%%%%%%%%%%%%%%%%%%%%5
\begin{align}\label{Fact.Sec.02.27}
S_{\alpha \beta}&=R_{\alpha \mu}R^{\mu}_{\beta}-\frac{1}{3}RR_{\alpha \beta}-\frac{1}{4}g_{\alpha \beta}
\left(R_{\mu \nu}R^{\mu \nu}-\frac{1}{3}R^{2}\right)
\nonumber
\\
&-\frac{1}{2}\left(\nabla _{\mu}\nabla _{\beta}R^{\mu}_{\alpha}
+\nabla _{\mu}\nabla _{\alpha}R^{\mu}_{\beta}\right)
+\frac{1}{3}\nabla _{\alpha}\nabla _{\beta}R+\frac{1}{2}\square R_{\alpha \beta}
-\frac{1}{12}g_{\alpha \beta}\square R
\end{align}
%%%%%%%%%%%%%%%%%%%%%%%%%%%%%%%%%%%%%%%%%%%%%%%%%%%%%%%%%%%%%%%%%%%%%%%%%%%%%%%%
Note that the tensor $S_{\mu \nu}$ is transverse and traceless along with all the other terms containing $\chi ^{\mu}_{\nu}$ and $C^{\mu}_{\nu}$, thanks to \eq{Fact.Sec.02.15}. In the expression for $\Sigma ^{(2)\mu}_{\nu}$, $t^{\mu}_{\nu}$ is an arbitrary integration constant, just like $\chi ^{\mu}_{\nu}$ in the first order. This tensor satisfies the following properties:
%%%%%%%%%%%%%%%%%%%%%%%%%%%%%%%%%%%%%%%%%%%%%%%%%%%%%%%%%%%%%%%%%5
\begin{equation}\label{Fact.Sec.02.28}
t^{\alpha}_{\alpha}=0,\qquad \nabla _{\alpha}t^{\alpha}_{\mu}=0.
\end{equation}
%%%%%%%%%%%%%%%%%%%%%%%%%%%%%%%%%%%%%%%%%%%%%%%%%%%%%%%%%%%%%%%%%%%
The traceless nature of $t_{\mu \nu}$ follows from the fact that $\Sigma ^{(2)\mu}_{\nu}$ is also traceless. Then, we can obtain the extrinsic curvature at second order as $K^{(2)\alpha}_{\beta}=\Sigma ^{(2)\alpha}_{\beta}+(1/4)\delta ^{\mu}_{\nu}K^{(2)}$. Thus the second-order correction to $h_{\mu \nu}$ can be obtained from the differential equation,
%%%%%%%%%%%%%%%%%%%%%%%%%%%%%%%%%%%%%%%%%%%%%%%%%%%%%%%%%%%%%%%%%%%%%%%%%%%%%%%%
\begin{align}\label{Fact.Sec.02.29}
-\frac{1}{2}\partial _{y}h^{(2)}_{\alpha \beta}&=-\frac{\ell ^{2}y}{2a^{4}}S_{\alpha \beta}+\frac{\ell ^{3}}{a^{4}}t_{\alpha \beta}+\frac{\ell ^{3}}{32a^{4}}g_{\alpha \beta}\left(R_{\mu \nu}R^{\mu \nu}-\frac{2}{9}R^{2}\right)
\nonumber
\\
&+\frac{\ell ^{2}}{4a^{6}}\Big[\frac{1}{2}\nabla _{\mu}\nabla _{\beta}\chi ^{\mu}_{\alpha }+\frac{1}{2}\nabla _{\mu}\nabla _{\alpha}\chi ^{\mu}_{\beta}-\frac{1}{2}\square \chi _{\alpha \beta}\Big]
-\frac{\ell}{16a^{8}}\Big[\chi _{\alpha \mu}\chi ^{\mu}_{\beta}-\frac{1}{4}g_{\alpha \beta}\chi ^{\mu \nu}\chi _{\mu \nu}\Big]
\nonumber
\\
&+\frac{\ell}{2a^{2}}\Big[\frac{1}{2}\nabla _{\mu}\nabla _{\beta}C^{\mu}_{\alpha }+\frac{1}{2}\nabla _{\mu}\nabla _{\alpha}C^{\mu}_{\beta}-\frac{1}{2}\nabla _{\alpha}\nabla _{\beta}C^{\mu}_{\mu}-\frac{1}{2}\square C_{\alpha \beta}+\frac{1}{\ell ^{2}}C_{\alpha \mu}C^{\mu}_{\beta}
\nonumber
\\
&-\frac{1}{4}g_{\alpha \beta}\left(\frac{1}{\ell ^{2}}C^{\mu \nu}C_{\mu \nu}+\nabla _{\mu}\nabla _{\nu}C^{\mu \nu}-\square C^{\mu}_{\mu}\right)\Big]
\nonumber
\\
&+\frac{1}{4}g_{\alpha \beta}\Big[\frac{5\ell}{24a^{8}}\chi ^{\mu \nu}\chi _{\mu \nu}+\frac{\ell}{6a^{2}}\left(\frac{1}{\ell ^{2}}C_{\mu \nu}C^{\mu \nu}+\nabla _{\mu}\nabla _{\alpha}C^{\mu \alpha}-\square C^{\mu}_{\mu}\right)\Big].
\end{align}
%%%%%%%%%%%%%%%%%%%%%%%%%%%%%%%%%%%%%%%%%%%%%%%%%%%%%%%%%%%%%%%%%%%%%%%%%%%%%%%%%%%%
This expression can be integrated to obtain the second-order correction to the metric as
%%%%%%%%%%%%%%%%%%%%%%%%%%%%%%%%%%%%%%%%%%%%%%%%%%%%%%%%%%%%%%%%%%%%%%%%%%%%%%%%%%%%
\begin{align}
q_{\alpha \beta}&=\left(\frac{\ell ^{3}y}{4a^{4}}-\frac{\ell ^{4}}{16a^{4}}\right)S_{\mu \nu}
-\frac{\ell ^{4}}{2a^{4}}t_{\mu \nu}(x)-\frac{\ell ^{4}}{64a^{4}}g_{\mu \nu}
\left(R_{\alpha \beta}R^{\alpha \beta}-\frac{1}{3}R^{2}\right)+B_{\mu \nu}(x)
\nonumber
\\
&-\frac{\ell ^{3}}{12a^{6}}\Big[\frac{1}{2}\nabla _{\mu}\nabla _{\beta}\chi ^{\mu}_{\alpha }+\frac{1}{2}\nabla _{\mu}\nabla _{\alpha}\chi ^{\mu}_{\beta}-\frac{1}{2}\square \chi _{\alpha \beta}\Big]+\frac{\ell ^{2}}{64a^{8}}\Big[\chi _{\alpha \mu}\chi ^{\mu}_{\beta}-\frac{1}{4}g_{\alpha \beta}\chi ^{\mu \nu}\chi _{\mu \nu}\Big]
\nonumber
\\
&-\frac{\ell ^{2}}{2a^{2}}\Big[\frac{1}{2}\nabla _{\mu}\nabla _{\beta}C^{\mu}_{\alpha }+\frac{1}{2}\nabla _{\mu}\nabla _{\alpha}C^{\mu}_{\beta}-\frac{1}{2}\nabla _{\alpha}\nabla _{\beta}C^{\mu}_{\mu}-\frac{1}{2}\square C_{\alpha \beta}+\frac{1}{\ell ^{2}}C_{\alpha \mu}C^{\mu}_{\beta}
\nonumber
\\
&-\frac{1}{4}g_{\alpha \beta}\left(\frac{1}{\ell ^{2}}C^{\mu \nu}C_{\mu \nu}+\nabla _{\mu}\nabla _{\nu}C^{\mu \nu}-\square C^{\mu}_{\mu}\right)\Big]
\nonumber
\\
&-\frac{1}{4}g_{\alpha \beta}\Big[\frac{5\ell ^{2}}{96a^{8}}\chi ^{\mu \nu}\chi _{\mu \nu}+\frac{\ell ^{2}}{6a^{2}}\left(\frac{1}{\ell ^{2}}C_{\mu \nu}C^{\mu \nu}+\nabla _{\mu}\nabla _{\alpha}C^{\mu \alpha}-\square C^{\mu}_{\mu}\right)\Big].
\end{align}
%%%%%%%%%%%%%%%%%%%%%%%%%%%%%%%%%%%%%%%%%%%%%%%%%%%%%%%%%%%%%%%%%%%%%%%%%%%%%%5
We can now add these zeroth-order, first-order and second-order corrections, in order to obtain the expression for $h_{\mu \nu}$ up to second-order as
%%%%%%%%%%%%%%%%%%%%%%%%%%%%%%%%%%%%%%%%%%%%%%%%%%%%%%%%%%%%%%%%%%%%%%%%%%%%%%%%5
\begin{align}\label{Fact.Sec.02.30}
h_{\mu \nu}&=a^{2}(y)\Big\lbrace g_{\mu \nu}-\frac{\ell ^{2}}{2a^{2}}\left(R_{\mu \nu}-\frac{1}{6}g_{\mu \nu}R\right)
-\frac{\ell}{2a^{4}}\chi _{\mu \nu}(x)+C_{\mu \nu}(x)
\nonumber
\\
&+\left(\frac{\ell ^{3}y}{4a^{4}}-\frac{\ell ^{4}}{16a^{4}}\right)S_{\mu \nu}
-\frac{\ell ^{4}}{2a^{4}}t_{\mu \nu}(x)-\frac{\ell ^{4}}{64a^{4}}g_{\mu \nu}
\left(R_{\alpha \beta}R^{\alpha \beta}-\frac{1}{3}R^{2}\right)+B_{\mu \nu}(x)
\nonumber
\\
&-\frac{\ell ^{3}}{12a^{6}}\Big[\frac{1}{2}\nabla _{\mu}\nabla _{\beta}\chi ^{\mu}_{\alpha }+\frac{1}{2}\nabla _{\mu}\nabla _{\alpha}\chi ^{\mu}_{\beta}-\frac{1}{2}\square \chi _{\alpha \beta}\Big]+\frac{\ell ^{2}}{64a^{8}}\Big[\chi _{\alpha \mu}\chi ^{\mu}_{\beta}-\frac{1}{4}g_{\alpha \beta}\chi ^{\mu \nu}\chi _{\mu \nu}\Big]
\nonumber
\\
&-\frac{\ell ^{2}}{2a^{2}}\Big[\frac{1}{2}\nabla _{\mu}\nabla _{\beta}C^{\mu}_{\alpha }+\frac{1}{2}\nabla _{\mu}\nabla _{\alpha}C^{\mu}_{\beta}-\frac{1}{2}\nabla _{\alpha}\nabla _{\beta}C^{\mu}_{\mu}-\frac{1}{2}\square C_{\alpha \beta}+\frac{1}{\ell ^{2}}C_{\alpha \mu}C^{\mu}_{\beta}
\nonumber
\\
&-\frac{1}{4}g_{\alpha \beta}\left(\frac{1}{\ell ^{2}}C^{\mu \nu}C_{\mu \nu}+\nabla _{\mu}\nabla _{\nu}C^{\mu \nu}-\square C^{\mu}_{\mu}\right)\Big]
\nonumber
\\
&-\frac{1}{4}g_{\alpha \beta}\Big[\frac{5\ell ^{2}}{96a^{8}}\chi ^{\mu \nu}\chi _{\mu \nu}+\frac{\ell ^{2}}{6a^{2}}\left(\frac{1}{\ell ^{2}}C_{\mu \nu}C^{\mu \nu}+\nabla _{\mu}\nabla _{\alpha}C^{\mu \alpha}-\square C^{\mu}_{\mu}\right)\Big]
\Big\rbrace,
\end{align}
%%%%%%%%%%%%%%%%%%%%%%%%%%%%%%%%%%%%%%%%%%%%%%%%%%%%%%%%%%%%%%%%%%%%%%%%%%%%%%%%%
where $B_{\mu \nu}$ is again a constant of integration. Thus, having obtained the metric $h_{\mu \nu}$ which includes corrections up to second order, we now calculate the effective action constructed out of it. 

As an aside, we would like to point out that the same procedure can be applied, in principle, to any arbitrary order in this gradient expansion scheme. Below we summarize the key steps of this procedure: (i) Given a solution correct up to $(n-1)$th order in this perturbative scheme, we first need to calculate the Ricci tensor $R^{(n)}_{\alpha \beta}$ and Ricci scalar $R^{(n)}$ at the $n$th order. (ii) Then, we need to use the Ricci scalar at $n$th order and $\Sigma ^{\alpha}_{\beta}$ and $K$ at lower orders to obtain $K^{(n)}$. (iii) We then have to integrate over the extra coordinate the differential equation for $\Sigma _{\alpha \beta}^{(n)}$ in order to get $\sigma _{\alpha \beta}$ at $n$th order. (d) Finally, we have to construct $K^{(n)}_{\alpha \beta}$ and integrate over the extra coordinate in order to obtain the metric corrected up to $n$th order.

It should be noted that second-order corrections were calculated in \cite{Kanno2002b} but with two assumptions: (i) the brane positions were fixed and (ii) quadratic terms in $\chi ^{\mu}_{\nu}$ could be neglected. However, in this work, we have kept our analysis completely general by relaxing both the assumptions; i.e., branes are not assumed to be fixed and quadratic corrections to $\chi ^{\mu}_{\nu}$ and $C_{\mu}^{\nu}$ terms are kept. 
%%%%%%%%%%%%%%%%%%%%%%%%%%%%%%%%%%%%%%%%%%%%%%%%%%%%%%%%%%%%%%%%%%%%%%%%%%%%%%%%%%%%%%%%%%%%%%%%%%%
%%%%%%%%%%%%%%%%%%%%%%%%%%%%%%%%%%%%%%%%%%%%%%%%%%%%%%%%%%%%%%%%%%%%%%%%%%%%%%%%%%%%%%%%%%%%%%%%%%%
%%%%%%%%%%%%%%%%%%%%%%%%%%%%%%%%%%%%%%%%%%%%%%%%%%%%%%%%%%%%%%%%%%%%%%%%%%%%%%%%%%%%%%%%%%%%%%%555%
\section{Effective Action}\label{Fact.Sec.03}

In this section we will determine the four-dimensional effective action corrected up to second order in the gradient expansion scheme. For that, we need the following pieces: (i) the bulk action $S_{\rm{bulk}}$, (ii) action for each of the branes represented by $S_{\pm}$, and finally (iii) the Gibbons-Hawking counterterm $S_{\rm{GH}}$. Note that in \cite{Kanno2005} the effective action was derived up to first order in the perturbative expansion to validate the metric factorizability. However, in this work, we generalize the analysis to second order in the gradient expansion scheme with variable brane positions. From the final structure, it becomes clear that the metric factorizability should hold at all orders in the gradient expansion scheme.

In order to determine the effective action we need to evaluate the determinant of $h_{\mu \nu}$. For this purpose, we will use \eq{Fact.Sec.02.07} and the following expression for the determinant:
%%%%%%%%%%%%%%%%%%%%%%%%%%%%%%%%%%%%%%%%%%%%%%%%%%%%%%%%%%%%%%%%%%%%%%%%%%%%%%%%%%%%%%%
\begin{align}\label{Fact.Sec.03.01}
h&=\frac{1}{24}\epsilon ^{\alpha \beta \mu \nu}\epsilon ^{\gamma \delta \rho \sigma}h_{\alpha \gamma} 
h_{\beta \delta} h_{\mu \rho} h_{\nu \sigma}
\nonumber
\\
&=\frac{a^{8}}{24}\epsilon ^{\alpha \beta \mu \nu}\epsilon ^{\gamma \delta \rho \sigma}
\left(g_{\alpha \gamma}+f_{\alpha \gamma}+q_{\alpha \gamma}\right)\left(g_{\beta \delta}+f_{\beta \delta}
+q_{\beta \delta}\right)\left(g_{\mu \rho}+f_{\mu \rho}+q_{\mu \rho}\right)
\left(g_{\nu \sigma}+f_{\nu \sigma}+q_{\nu \sigma}\right)
\nonumber
\\
&=\frac{a^{8}}{24}\epsilon ^{\alpha \beta \mu \nu}\epsilon ^{\gamma \delta \rho \sigma}
\left(g_{\alpha \gamma} g_{\beta \delta} g_{\mu \rho} g_{\nu \sigma}
+4f_{\alpha \gamma} g_{\beta \delta} g_{\mu \rho} g_{\nu \sigma}
+6f_{\alpha \gamma} f_{\beta \delta} g_{\mu \rho} g_{\nu \sigma}
+4q_{\alpha \gamma} g_{\beta \delta} g_{\mu \rho} g_{\nu \sigma}\right)
\nonumber
\\
&=\frac{a^{8}}{24}\Big[\epsilon ^{\alpha \beta \mu \nu}\epsilon ^{\gamma \delta \rho \sigma}
g_{\alpha \gamma} g_{\beta \delta} g_{\mu \rho} g_{\nu \sigma}
-4f^{\alpha}_{\beta}\epsilon _{\alpha \gamma \mu \nu}\epsilon ^{\beta \gamma \mu \nu}
-6f^{\alpha}_{\beta}f^{\gamma}_{\delta}\epsilon _{\alpha \gamma \mu \nu}\epsilon ^{\beta \delta \mu \nu}
-4q^{\alpha}_{\beta}\epsilon _{\alpha \gamma \mu \nu}\epsilon ^{\beta \gamma \mu \nu}\Big]
\nonumber
\\
&=\frac{a^{8}}{24}g\left[24+24f^{\mu}_{\mu}+24q^{\alpha}_{\alpha}+12\left(f^{\mu}_{\mu}f^{\alpha}_{\alpha}
-f_{\mu \nu}f^{\mu \nu}\right) \right]
\nonumber
\\
&=a^{8}g\Big[1-\frac{\ell ^{2}}{6a^{2}}R-\frac{\ell ^{4}}{16a^{4}}\left(3R_{\alpha \beta}R^{\alpha \beta}
-R^{2}\right)-\frac{17 \ell ^{2}}{96a^{8}}\chi _{\mu \nu}\chi ^{\mu \nu}
\nonumber
\\
&-\frac{\ell ^{2}}{6a^{2}}\left(\frac{1}{\ell ^{2}}C_{\mu \nu}C^{\mu \nu}+\nabla _{\mu}\nabla _{\alpha}C^{\mu \alpha}-\square C^{\mu}_{\mu}\right)+\left(C+B+\frac{1}{2}C^{2}-\frac{1}{2}C_{\mu nu}C^{\mu \nu}\right)\Big],
\end{align}
%%%%%%%%%%%%%%%%%%%%%%%%%%%%%%%%%%%%%%%%%%%%%%%%%%%%%%%%%%%%%%%%%%%%%%%%%%%%%%%%%%%%%%%%%%%
where we have used the following identities:
%%%%%%%%%%%%%%%%%%%%%%%%%%%%%%%%%%%%%%%%%%%%%%%%%%%%%%%%%%%%%%%%%%%%%%%%%%%%%%%%%%%%55
\begin{align}
\epsilon ^{\alpha \beta \mu \nu}&=-\epsilon _{\alpha \beta \mu \nu}
\label{Fact.Sec.03.02}
\\
\epsilon ^{\alpha \beta \mu \nu}\epsilon _{\gamma \beta \mu \nu}&=-6\delta ^{\alpha}_{\gamma}
\label{Fact.Sec.03.03}
\\
\epsilon ^{\alpha \beta \mu \nu}\epsilon _{\gamma \rho \mu \nu}&=-2\left(\delta ^{\alpha}_{\gamma} 
\delta ^{\beta} _{\rho}-\delta ^{\alpha}_{\rho}\delta ^{\beta}_{\gamma}\right).
\label{Fact.Sec.03.04}
\end{align}
%%%%%%%%%%%%%%%%%%%%%%%%%%%%%%%%%%%%%%%%%%%%%%%%%%%%%%%%%%%%%%%%%%%%%%%%%%%%%%%%%%%%%%
Then, we obtain
%%%%%%%%%%%%%%%%%%%%%%%%%%%%%%%%%%%%%%%%%%%%%%%%%%%%%%%%%%%%%%%%%%%%%%%%%%%%%%%%%%%%%%%%
\begin{align}\label{Fact.Sec.03.05}
\sqrt{-\mathcal{G}}=\sqrt{-h}&=a^{4}\sqrt{-g}\Big[1-\frac{\ell ^{2}}{6a^{2}}R-\frac{\ell ^{4}}{16a^{4}}\left(3R_{\alpha \beta}R^{\alpha \beta}
-R^{2}\right)-\frac{17 \ell ^{2}}{96a^{8}}\chi _{\mu \nu}\chi ^{\mu \nu}
\nonumber
\\
&-\frac{\ell ^{2}}{6a^{2}}\left(\frac{1}{\ell ^{2}}C_{\mu \nu}C^{\mu \nu}+\nabla _{\mu}\nabla _{\alpha}C^{\mu \alpha}-\square C^{\mu}_{\mu}\right)+\left(C+B+\frac{1}{2}C^{2}-\frac{1}{2}C_{\mu nu}C^{\mu \nu}\right)\Big]^{1/2}
\nonumber
\\
&=a^{4}\sqrt{-g}\Big[1-\frac{\ell ^{2}}{12a^{2}}R
-\frac{\ell ^{4}}{32a^{4}}\left(3R_{\alpha \beta}R^{\alpha \beta}-\frac{8}{9}R^{2}\right)-\frac{17 \ell ^{2}}{192a^{8}}\chi _{\mu \nu}\chi ^{\mu \nu}
\nonumber
\\
&-\frac{\ell ^{2}}{12a^{2}}\left(\frac{1}{\ell ^{2}}C_{\mu \nu}C^{\mu \nu}+\nabla _{\mu}\nabla _{\alpha}C^{\mu \alpha}-\square C^{\mu}_{\mu}\right)\Big]\Big\lbrace 1+\frac{1}{2}\left(C+B+\frac{1}{2}C^{2}-\frac{1}{2}C_{\mu nu}C^{\mu \nu}\right)\Big\rbrace.
\end{align}
%%%%%%%%%%%%%%%%%%%%%%%%%%%%%%%%%%%%%%%%%%%%%%%%%%%%%%%%%%%%%%%%%%%%%%%%%%%%%%%%%%%%%%%%%%%
Note that all the terms on the last line in the second bracket do not have an effect on the effective action. Then, following \cite{Kanno2005}, we could neglect this term. However, there are two crucial differences from the analysis presented in \cite{Kanno2005}: (i) We have incorporated second-order corrections to the effective action, while in \cite{Kanno2005} only first-order corrections were considered and secondly (b) we have kept both the integration constants $\chi _{\mu \nu}$ and $C_{\mu \nu}$ in contrast to \cite{Kanno2005}. 

Having obtained the bulk metric, it is now trivial to calculate the bulk action, with second-order correction terms included. For that purpose we substitute the determinant $\sqrt{-h}$ which includes second-order corrections, to the bulk action. With this factor included in the bulk action we arrive at
%%%%%%%%%%%%%%%%%%%%%%%%%%%%%%%%%%%%%%%%%%%%%%%%%%%
\begin{align}\label{Fact.Sec.03.06}
S_{\rm{bulk}}&=\frac{1}{2\kappa ^{2}}\int d^{5}x\sqrt{-\mathcal{G}}\left[\mathcal{R}+\frac{12}{\ell ^{2}}\right]
\nonumber
\\
&=-\frac{8}{\kappa ^{2}\ell ^{2}}\int d^{4}x\sqrt{-g}\Big[\frac{\ell}{4}\left(a_{+}^{4}-a_{-}^{4}\right)
-\frac{\ell ^{3}R}{24}\left(a_{+}^{2}-a_{-}^{2}\right)+\frac{\ell ^{4}}{32}\left(\phi _{+}-\phi _{-}\right)
\left(3R^{\alpha}_{\beta}R^{\beta}_{\alpha}-\frac{8}{9}R^{2}\right)
\nonumber
\\
&+\frac{17 \ell ^{2}}{768}\left(\frac{1}{a_{+}^{4}}-\frac{1}{a_{-}^{4}}\right)\chi _{\mu \nu}\chi ^{\mu \nu}
-\frac{\ell ^{3}}{24}\left(a_{+}^{2}-a_{-}^{2}\right)\left(\frac{1}{\ell ^{2}}C_{\mu \nu}C^{\mu \nu}+\nabla _{\mu}\nabla _{\alpha}C^{\mu \alpha}-\square C^{\mu}_{\mu}\right)\Big],
\end{align}
%%%%%%%%%%%%%%%%%%%%%%%%%%%%%%%%%%%%%%%%%%%%%%%%%%
where $\phi _{+}$ and $\phi _{-}$ are the respective brane positions defined through \eq{Fact.Sec.02.02}. We have defined the warp factor $a^{2}$ at the position of the branes $\phi _{+}$ and $\phi _{-}$ as $a^{2}_{+}$ and $a^{2}_{-}$, respectively. In order to arrive at the second line, we have used the result for bulk Ricci scalar as $\mathcal{R}=-20/\ell ^{2}$. The next thing to calculate is the action corresponding to the brane tension. For this we require the induced metric on each brane, with the following expression:
%%%%%%%%%%%%%%%%%%%%%%%%%%%%%%%%%%%%%%%%%%%%%%%%%%%%%%%%%%
\begin{align}\label{Fact.Sec.03.07}
g_{\alpha \beta}^{\pm}\left(y=\phi _{\pm},x\right)=a_{\pm}^{2}\left[g_{\alpha \beta}(x)
+f_{\alpha \beta}\left(\phi _{\pm},x\right)+q_{\alpha \beta}\left(\phi _{\pm},x\right)\right]
+\partial _{\alpha}\phi _{\pm}\partial _{\beta}\phi _{\pm}.
\end{align}
%%%%%%%%%%%%%%%%%%%%%%%%%%%%%%%%%%%%%%%%%%%%%%%%%%%%%%%%%%%%%%%
Then, the determinant of the induced metric turns out to have the following expression:
%%%%%%%%%%%%%%%%%%%%%%%%%%%%%%%%%%%%%%%%%%%%%%%%%%%%%%%%%%%%%%
\begin{align}\label{Fact.Sec.03.08}
\sqrt{-g_{\pm}}&=a_{\pm}^{4}\sqrt{-g}
\Big[1+\frac{1}{a_{\pm}^{2}}\partial _{\mu}\phi _{\pm}\partial ^{\mu}\phi _{\pm}
-\frac{\ell ^{2}}{6a_{\pm}^{2}}R-\frac{\ell ^{4}}{16a_{\pm}^{4}}\left(3R_{\alpha \beta}R^{\alpha \beta}
-R^{2}\right)-\frac{17 \ell ^{2}}{96a_{\pm}^{8}}\chi _{\mu \nu}\chi ^{\mu \nu}
\nonumber
\\
&-\frac{\ell ^{2}}{6a_{\pm}^{2}}\left(\frac{1}{\ell ^{2}}C_{\mu \nu}C^{\mu \nu}+\nabla _{\mu}\nabla _{\alpha}C^{\mu \alpha}-\square C^{\mu}_{\mu}\right)\Big]^{1/2}
\nonumber
\\
&=a_{\pm}^{4}\sqrt{-g}
\Big[1+\frac{1}{2a_{\pm}^{2}}\partial _{\mu}\phi _{\pm}\partial ^{\mu}\phi _{\pm}
-\frac{\ell ^{2}}{12a_{\pm}^{2}}R-\frac{\ell ^{4}}{32a_{\pm}^{4}}\left(3R_{\alpha \beta}R^{\alpha \beta}
-\frac{8}{9}R^{2}\right)-\frac{17 \ell ^{2}}{192a_{\pm}^{8}}\chi _{\mu \nu}\chi ^{\mu \nu}
\nonumber
\\
&-\frac{\ell ^{2}}{12a_{\pm}^{2}}\left(\frac{1}{\ell ^{2}}C_{\mu \nu}C^{\mu \nu}+\nabla _{\mu}\nabla _{\alpha}C^{\mu \alpha}-\square C^{\mu}_{\mu}\right)\Big].
\end{align}
%%%%%%%%%%%%%%%%%%%%%%%%%%%%%%%%%%%%%%%%%%%%%%%%%%%%%%%%%%%%%%%%%%%%%
With the help of the above equation, the action on the two branes can be written as
%%%%%%%%%%%%%%%%%%%%%%%%%%%%%%%%%%%%%%%%%%%%%%%%%%%%%%%%
\begin{align}\label{Fact.Sec.03.09}
S_{\pm}&=\mp \frac{6}{\kappa ^{2}\ell}\int d^{4}x \sqrt{-g_{\pm}}
\nonumber
\\
&=\mp \frac{6}{\kappa ^{2}\ell}\int d^{4}x\sqrt{-g}\Big[a^{4}_{\pm}
+\frac{1}{2}a^{2}_{\pm}\left(\partial _{\mu} \phi _{\pm}\partial ^{\mu}\phi _{\pm}\right)
-\frac{\ell ^{2}}{12}Ra^{2}_{\pm}-\frac{\ell ^{4}}{32}
\left(3R^{\alpha}_{\beta}R^{\beta}_{\alpha}-\frac{8}{9}R^{2}\right)-\frac{17 \ell ^{2}}{192a_{\pm}^{4}}\chi _{\mu \nu}\chi ^{\mu \nu}
\nonumber
\\
&-\frac{\ell ^{2}}{12}a_{\pm}^{2}\left(\frac{1}{\ell ^{2}}C_{\mu \nu}C^{\mu \nu}+\nabla _{\mu}\nabla _{\alpha}C^{\mu \alpha}-\square C^{\mu}_{\mu}\right)\Big].
\end{align}
%%%%%%%%%%%%%%%%%%%%%%%%%%%%%%%%%%%%%%%%%%%%%%%%%%%%%%%%%%%%%5
Then, simple addition of the two actions $S_{+}$ and $S_{-}$ leads to
%%%%%%%%%%%%%%%%%%%%%%%%%%%%%%%%%%%%%%%%%%%%%%%%%%%%%%%%%
\begin{align}\label{Fact.Sec.03.10}
S_{+}+S_{-}&=-\frac{6}{\kappa ^{2}\ell}\int d^{4}x\sqrt{-g}
\Big[\left(a^{4}_{+}-a^{4}_{-}\right)+\frac{1}{2}\left(a^{2}_{+}\partial _{\mu}\phi _{+}\partial ^{\mu}\phi _{+}
-a^{2}_{-}\partial _{\mu}\phi _{-}\partial ^{\mu}\phi _{-}\right)
-\frac{\ell ^{2}}{12}R\left(a_{+}^{2}-a_{-}^{2}\right)
\nonumber
\\
&-\frac{17 \ell ^{2}}{192}\left(\frac{1}{a_{+}^{4}}-\frac{1}{a_{-}^{4}}\right)\chi _{\mu \nu}\chi ^{\mu \nu}-\frac{\ell ^{2}}{12}\left(a_{+}^{2}-a_{-}^{2}\right)\left(\frac{1}{\ell ^{2}}C_{\mu \nu}C^{\mu \nu}+\nabla _{\mu}\nabla _{\alpha}C^{\mu \alpha}-\square C^{\mu}_{\mu}\right)\Big].
\end{align}
%%%%%%%%%%%%%%%%%%%%%%%%%%%%%%%%%%%%%%%%%%%%%%%%%%%%%%%%%%%%%5
Finally, we need to calculate the counterterm provided by Gibbons and Hawking. For that we need to calculate the extrinsic curvature or, more importantly, its trace. The extrinsic curvature is defined as \cite{Kanno2005}
%%%%%%%%%%%%%%%%%%%%%%%%%%%%%%%%%%%%%%%%%%%%%%%%%%%%%%%%%%%%%%
\begin{align}
\mathcal{K}_{\mu \nu}=n_{a}\left[\left(\frac{\partial ^{2}x^{a}}{\partial \xi ^{\mu}\xi ^{\nu}}\right)+\Gamma ^{a}_{bc}\frac{\partial x^{b}}{\partial \xi ^{\mu}}\frac{\partial x^{c}}{\partial \xi ^{\nu}}\right],
\end{align}
%%%%%%%%%%%%%%%%%%%%%%%%%%%%%%%%%%%%%%%%%%%%%%%%%%%%%%%%%%%%%%
where $n_{a}$ is the vector normal to the brane and the required Christoffel symbols are
%%%%%%%%%%%%%%%%%%%%%%%%%%%%%%%%%%%%%%%%%%%%%%%%%%%%%%%%%%%%%%
\begin{align}
\Gamma ^{y}_{\mu \nu}&=\frac{a^{2}}{\ell}\left(g_{\mu \nu}+f_{\mu \nu}+q_{\mu \nu}\right)-\frac{a^{2}}{2}\left(\partial _{y}f_{\mu \nu}+\partial _{y}q_{\mu \nu}\right)
\\
\Gamma ^{\alpha}_{y\mu}&=-\frac{1}{\ell}\delta ^{\alpha}_{\mu}+\frac{1}{2}g^{\alpha \beta}\left(\partial _{y}f_{\alpha \beta}+\partial _{y}q_{\alpha \beta}\right).
\end{align}
%%%%%%%%%%%%%%%%%%%%%%%%%%%%%%%%%%%%%%%%%%%%%%%%%%%%%%%%%%%%%%%
Then, the extrinsic curvature turns out to be
%%%%%%%%%%%%%%%%%%%%%%%%%%%%%%%%%%%%%%%%%%%%%%%%%%%%%%%%%%%%%%%
\begin{align}
\mathcal{K}^{\pm}_{\mu \nu}&=n_{y}\left[\nabla _{\mu}\nabla _{\nu}\phi _{\pm}+\frac{2}{\ell}\partial _{\mu}\phi _{\pm}\partial _{\nu}\phi _{\pm}+\frac{a_{\pm}^{2}}{\ell}\left(g_{\mu \nu}+f_{\mu \nu}+q_{\mu \nu}-\frac{\ell}{2}\partial _{y}f_{\mu \nu}-\frac{\ell}{2}\partial _{y}q_{\mu \nu}\right)\right],
\end{align}
%%%%%%%%%%%%%%%%%%%%%%%%%%%%%%%%%%%%%%%%%%%%%%%%%%%%%%%%%%%%%%%%5
the trace of which has the following expression:
%%%%%%%%%%%%%%%%%%%%%%%%%%%%%%%%%%%%%%%%%%%%%%%%%%%%%%%%%%%%%%%%%%%%%
\begin{align}\label{Fact.Sec.03.11}
\mathcal{K}_{\pm}&=n_{y}\Big[\frac{4}{\ell}+\frac{1}{a_{\pm}^{2}}\square \phi _{\pm}+
\frac{1}{\ell a_{\pm}^{2}}\partial _{\mu}\phi _{\pm}\partial ^{\mu}\phi _{\pm}
+\frac{\ell}{6a_{\pm}^{2}}R
-\frac{\ell ^{3}}{8a_{\pm}^{4}}\left(R_{\mu \nu}R^{\mu \nu}-\frac{1}{9}R^{2}\right)
\nonumber
\\
&-\frac{7 \ell}{24a_{\pm}^{8}}\chi _{\mu \nu}\chi ^{\mu \nu}+\frac{\ell}{6a_{\pm}^{2}}\left(\frac{1}{\ell ^{2}}C_{\mu \nu}C^{\mu \nu}+\nabla _{\mu}\nabla _{\alpha}C^{\mu \alpha}-\square C^{\mu}_{\mu}\right)\Big].
\end{align}
%%%%%%%%%%%%%%%%%%%%%%%%%%%%%%%%%%%%%%%%%%%%%%%%%%%%%%%%%%%%%%%%%%%%%%%%%%%%%%%%%%%%%%
The action corresponding to the Gibbon-Hawking counterterm has the expression
%%%%%%%%%%%%%%%%%%%%%%%%%%%%%%%%%%%%%%%%%%%%%%%%%%%%%%%%%%%%%
\begin{align}\label{Fact.Sec.03.12}
S_{\rm{GH}}&=\frac{2}{\kappa ^{2}}\int d^{4}x\sqrt{-g_{+}}\mathcal{K}_{+}
-\frac{2}{\kappa ^{2}}\int d^{4}x\sqrt{-g_{-}}\mathcal{K}_{-}
\nonumber
\\
&=\frac{2}{\kappa ^{2}}\int d^{4}x\sqrt{-g}\Big[\frac{4}{\ell}\left(a^{4}_{+}-a^{4}_{-}\right)
-\frac{\ell}{6}R\left(a_{+}^{2}-a_{-}^{2}\right)+\frac{3}{\ell}\left(a^{2}_{+}\partial _{\mu}\phi _{+}\partial ^{\mu}\phi _{+}
-a^{2}_{-}\partial _{\mu}\phi _{-}\partial ^{\mu}\phi _{-}\right)
\nonumber
\\
&-\frac{31\ell}{48}\left(\frac{1}{a_{+}^{4}}-\frac{1}{a_{-}^{4}}\right)\chi _{\mu \nu}\chi ^{\mu \nu}-\frac{\ell}{6}\left(a_{+}^{2}-a_{-}^{2}\right)\left(\frac{1}{\ell ^{2}}C_{\mu \nu}C^{\mu \nu}+\nabla _{\mu}\nabla _{\alpha}C^{\mu \alpha}-\square C^{\mu}_{\mu}\right)\Big].
\end{align}
%%%%%%%%%%%%%%%%%%%%%%%%%%%%%%%%%%%%%%%%%%%%%%%%%%%%%%%%%%%%%%%%5
Thus, substitution of the bulk action, brane tension, and Gibbon-Hawking counterterm leads to the complete four-diemnsional effective action, which has the expression
%%%%%%%%%%%%%%%%%%%%%%%%%%%%%%%%%%%%%%%%%%%%%%%%%%%%%%%%
\begin{align}\label{Fact.Sec.03.13}
S_{\rm{tot}}&=S_{\rm{bulk}}+S_{+}+S_{-}+S_{\rm{GH}}
\nonumber
\\
&=\frac{\ell}{2\kappa ^{2}}\int d^{4}x\sqrt{-g}\Big[\left(a_{+}^{2}-a_{-}^{2}\right)R
+\frac{6}{\ell ^{2}}\left(a^{2}_{+}\partial _{\mu}\phi _{+}\partial ^{\mu}\phi _{+}
-a^{2}_{-}\partial _{\mu}\phi _{-}\partial ^{\mu}\phi _{-}\right)
\nonumber
\\
&-\frac{\ell}{2}\left(\phi _{+}-\phi _{-}\right)
\left(3R^{\alpha}_{\beta}R^{\beta}_{\alpha}-\frac{8}{9}R^{2}\right)-\frac{15}{8}\left(\frac{1}{a_{+}^{4}}-\frac{1}{a_{-}^{4}}\right)\chi _{\mu \nu}\chi ^{\mu \nu}
\nonumber
\\
&+\left(a_{+}^{2}-a_{-}^{2}\right)\left(\frac{1}{\ell ^{2}}C_{\mu \nu}C^{\mu \nu}+\nabla _{\mu}\nabla _{\alpha}C^{\mu \alpha}-\square C^{\mu}_{\mu}\right)\Big].
\end{align}
%%%%%%%%%%%%%%%%%%%%%%%%%%%%%%%%%%%%%%%%%%%%%%%%%%%%%%%%%
Note that if we had dropped all the second-order terms we would arrive at the result obtained in \cite{Kanno2005}. However, since we have worked with branes with variable position and kept terms up to second order this provides a direct generalization of the results obtained in \cite{Kanno2005}. The nice separation of terms into extra dimensional part and a brane part shows the validity of the factorizable metric ansatz up to second order (it had been shown only up to first order in \cite{Kanno2005}). Indeed, we could do more from the above action. The structure suggests that the third-order terms would be associated with $a^{-2}$, the fourth-order terms will be connected to $a^{-4}$ and so on. Thus the nth-order term would be associated with a $a^{-2(n-2)}$ term. These terms would be independent of the part that depends on the brane coordinates. Thus the effective action when third-order corrections are incorporated would contain terms like $R_{\alpha \mu}R^{\mu \nu}R_{\nu}^{\alpha}\times \left(a_{+}^{-2}-a_{-}^{-2}\right)$ and $\chi _{\alpha \mu}\chi ^{\mu \beta}\chi _{\beta}^{\alpha}(a_{+}^{-8}-a_{-}^{-8})$. All these terms will appear with the extra dimensional part separated from the terms dependent on the brane coordinates. Moreover, it should be noted that only the difference of the various powers of warp factors between the two branes enters the picture. From this, we could conclude that the effective action would be factorizable at all orders and only the difference between these moduli fields appears in the effective action. 

In order to understand the effective action in greater detail, we vary the action with respect to $g_{\mu \nu}$ with the assumption of a fixed brane; i.e., $\phi _{+}$ and $\phi _{-}$ are assumed to be independent of $x^{\mu}$. Then, the equation of motion in absence of any matter field obtained from arbitrary variation of $S_{\rm{tot}}$ with respect to $g_{\mu \nu}$ turns out to be (neglecting the $C^{\mu}_{\nu}$ terms)
%%%%%%%%%%%%%%%%%%%%%%%%%%%%%%%%%%%%%%%%%%%%%%%%%%%%%%%%%%%%%%%%%%%%%%%%%%%%%%%%%%%%
\begin{align}\label{Fact.Sec.03.14}
\frac{\ell}{2\kappa ^{2}}\left(a_{+}^{2}-a_{-}^{2}\right)&\times\left(R_{\mu \nu}-\frac{1}{2}g_{\mu \nu}R\right)
-\frac{\ell ^{2}}{4\kappa ^{2}}\left(\phi _{+}-\phi _{-}\right)
\Big\lbrace 6R_{\mu \alpha}R^{\alpha}_{\nu}-\frac{16}{9}RR_{\mu \nu}
-\frac{1}{2}g_{\mu \nu}\left(3R_{\alpha \beta}R^{\alpha \beta}-\frac{8}{9}R^{2}\right)
\nonumber
\\
&+\frac{16}{9}\nabla _{\mu}\nabla _{\nu}R
-6\nabla _{\alpha}\nabla _{\mu}R^{\alpha}_{\nu}+3\nabla _{\alpha}\nabla ^{\alpha}R_{\mu \nu}-\frac{1}{2}g_{\mu \nu}\left(3\nabla _{\alpha}\nabla _{\beta}R^{\alpha \beta}-\frac{16}{9}\square R\right)
\Big\rbrace
\nonumber
\\
&-\frac{15\ell }{16\kappa ^{2}}\left(\frac{1}{a_{+}^{4}}-\frac{1}{a_{-}^{4}}\right)\left\lbrace 2\chi _{\mu \alpha}\chi ^{\alpha}_{\nu}-\frac{1}{2}g_{\mu \nu}\chi ^{\alpha \beta}\chi _{\alpha \beta}\right\rbrace =0.
\end{align}
%%%%%%%%%%%%%%%%%%%%%%%%%%%%%%%%%%%%%%%%%%%%%%%%%%%%%%%%%%%%%%%%%%%%%%%%%%%%%%%%%%%%%%
However, this equation contains higher-order derivatives of the metric, and it has nonlocal terms originating from the tensor $\chi ^{\mu \nu}$. In order to avoid the appearance of any ghost field, these higher derivative terms must vanish along with some suitable choice for this field $\chi ^{\mu}_{\nu}$. For the Proper choice of $\chi ^{\mu}_{\nu}$, this condition yields the following equation:
%%%%%%%%%%%%%%%%%%%%%%%%%%%%%%%%%%%%%%%%%%%%%%%%%%%%%%%%%%%%%%%%%%%%%%%%%%%%%%%%%%%%%%%
\begin{align}\label{Fact.Sec.03.15}
2\nabla _{\mu}\nabla _{\nu}R
-6\nabla _{\alpha}\nabla _{\mu}R^{\alpha}_{\nu}+3\nabla _{\alpha}\nabla ^{\alpha}R_{\mu \nu}
-\frac{1}{2}g_{\mu \nu}\left(3\nabla _{\alpha}\nabla _{\beta}R^{\alpha \beta}-2\square R\right)=0.
\end{align}
%%%%%%%%%%%%%%%%%%%%%%%%%%%%%%%%%%%%%%%%%%%%%%%%%%%%%%%%%%%%%%%%%%%%%%%%%%%%%%%
The interesting aspect of this equation is that the trace part leads to $\nabla _{\mu}\nabla _{\nu}G^{\mu \nu}=0$, which is automatically satisfied by the Bianchi identity. Thus, the action $S_{\rm{tot}}$ with \eq{Fact.Sec.03.15} imposed represents a higher-order gravity theory. It would be interesting to investigate possible spherically symmetric solutions, solar system tests, and the nature of gravitational waves originating from this action. This is a work in progress and will be presented elsewhere.

%%%%%%%%%%%%%%%%%%%%%%%%%%%%%%%%%%%%%%%%%%%%%%%%%%%%%%%%%%%%%%%%%%%%%%%%%%%%%%%%%%%%%%%%%%%%%%%%%%%%%%%%
%%%%%%%%%%%%%%%%%%%%%%%%%%%%%%%%%%%%%%%%%%%%%%%%%%%%%%%%%%%%%%%%%%%%%%%%%%%%%%%%%%%%%%%%%%%%%%%%%%%%%%%%
%%%%%%%%%%%%%%%%%%%%%%%%%%%%%%%%%%%%%%%%%%%%%%%%%%%%%%%%%%%%%%%%%%%%%%%%%%%%%%%%%%%%%%%%%%%%%%%%%%%%%%%%
\section{Equivalence with Scalar-Tensor Gravity}\label{Fact.Sec.04}

In first order we have two arbitrary constants, $C_{\mu \nu}(x)$ and $\chi _{\mu \nu}(x)$, both of which are independent of the extra coordinate, and are dependent on the brane coordinates. Let us exploit these two tensors and obtain some simplified results. First, we can use $C_{\mu \nu}$ such that $f_{\mu \nu}(y=\phi _{+},x)=0$. This can be seen explicitly from \eq{Fact.Sec.02.16}, which under the above condition reduces to the following form:
%%%%%%%%%%%%%%%%%%%%%%%%%%%%%%%%%%%%%%%%%%%%%%%%%%%%%%%%%%%%%%%%%%%
\begin{equation}\label{Fact.Sec.04.01}
f_{\mu \nu}(y=\phi _{+},x)=-\frac{\ell ^{2}}{2a_{+}^{2}}\left(R_{\mu \nu}-\frac{1}{6}g_{\mu \nu}R\right)-\frac{\ell}{2a_{+}^{4}}\chi _{\mu \nu}(x)+C_{\mu \nu}(x)=0.
\end{equation}
%%%%%%%%%%%%%%%%%%%%%%%%%%%%%%%%%%%%%%%%%%%%%%%%%%%%%%%%%%%%%%%%%%%%%%%%%%
It may be noted that we cannot use $\chi _{\mu \nu}$ to set $f_{\mu \nu}(y=\phi _{-},x)=0$. In order to achieve this, we need the arbitrary constant $\chi _{\mu \nu}$ to satisfy 
%%%%%%%%%%%%%%%%%%%%%%%%%%%%%%%%%%%%%%%%%%%%%%%%%%%%%%%5%%%%%%%%%%%%%%5
\begin{equation}\label{Fact.Sec.04.02}
f_{\mu \nu}(y=\phi _{-},x)=\frac{\ell ^{2}}{2}\left(\frac{1}{a_{+}^{2}}-\frac{1}{a_{-}^{2}}\right)
\left(R_{\mu \nu}-\frac{1}{6}g_{\mu \nu}R\right)+\frac{\ell}{2}\left(\frac{1}{a_{+}^{4}}-\frac{1}{a_{-}^{4}}\right)\chi _{\mu \nu}(x)=0.
\end{equation}
%%%%%%%%%%%%%%%%%%%%%%%%%%%%%%%%%%%%%%%%%%%%%%%%%%%%%%%%%%%%%%%%%%%%%%%%%
which cannot be achieved due to the tracelessness of $\chi _{\mu \nu}$. Thus, rather than working along this line, we can impose another (single) boundary condition, $a^{2}_{+}f_{\mu \nu}(y=\phi _{+},x)=a^{2}_{-}f_{\mu \nu}(y=\phi _{-},x)$. The equation satisfied by $\chi _{\mu \nu}$ now turns out to be
%%%%%%%%%%%%%%%%%%%%%%%%%%%%%%%%%%%%%%%%%%%%%%%%%%%%%%%%%%%%%%%%%%%%%%%%%
\begin{equation}
\left(a_{+}^{2}-a_{-}^{2}\right)C_{\mu \nu}=\frac{\ell}{2}\left(\frac{1}{a_{+}^{2}}-\frac{1}{a_{-}^{2}}\right)\chi _{\mu \nu}.
\end{equation}
%%%%%%%%%%%%%%%%%%%%%%%%%%%%%%%%%%%%%%%%%%%%%%%%%%%%%%%%%%%%%%%%%%%%%%%%%
which can always be satisfied by properly choosing the arbitrary tensor $C_{\mu \nu}$ to be traceless. 

Then, in a similar manner, we can use $t_{\mu \nu}$ and $B_{\mu \nu}$ to set $a^{2}_{+}q_{\mu \nu}(y=\phi _{+},x)=a^{2}_{-}q_{\mu \nu}(y=\phi _{-},x)$ such that (ignoring the arbitrary tensors $\chi _{\mu \nu}$ and $C_{\mu \nu}$)
%%%%%%%%%%%%%%%%%%%%%%%%%%%%%%%%%%%%%%%%%%%%%%%%%%%%%%%%%%%%%%%%%%%%%%%%%
\begin{align}
S_{\mu \nu} &\times \left[\left(\frac{\ell ^{3}y}{4a_{+}^{2}}-\frac{\ell ^{4}}{16a_{+}^{2}}\right)-\left(\frac{\ell ^{3}y}{4a_{-}^{2}}-\frac{\ell ^{4}}{16a_{-}^{2}}\right)\right]
-\left(\frac{\ell ^{4}}{2a^{2}_{+}}-\frac{\ell ^{4}}{2a^{2}_{-}}\right)t_{\mu \nu}(x)
\nonumber
\\
&=\left(\frac{\ell ^{4}}{64a_{+}^{2}}-\frac{\ell ^{4}}{64a_{-}^{2}}\right)g_{\mu \nu}
\left(R_{\alpha \beta}R^{\alpha \beta}-\frac{1}{3}R^{2}\right)-\left(a_{+}^{2}-a_{-}^{2}\right)B_{\mu \nu}(x).
\end{align}
%%%%%%%%%%%%%%%%%%%%%%%%%%%%%%%%%%%%%%%%%%%%%%%%%%%%%%%%%%%%%%%%%%%%%%%%%%%
Note that the trace of the left-hand side vanishes. Thus, trace of arbitrary tensor $B_{\mu \nu}$ should such that the above equation is satisfied. The same argument holds at all orders. Thus, finally we have 
%%%%%%%%%%%%%%%%%%%%%%%%%%%%%%%%%%%%%%%%%%%%%%%%%%%%%%%%%%%%%%%%%%%%%%%%%%
\begin{align}
h_{\mu \nu}(y=\phi _{+},x)&=a^{2}\left(\phi _{+}\right)\left[g_{\mu \nu}+f_{\mu \nu}\left(\phi _{+},x\right)
+q_{\mu \nu}\left(\phi _{+},x\right)+\ldots \right]
\label{Fact.Sec.04.03a}
\\
h_{\mu \nu}(y=\phi _{-},x)&=a^{2}\left(\phi _{-}\right)\left[g_{\mu \nu}+f_{\mu \nu}\left(\phi _{-},x\right)
+q_{\mu \nu}\left(\phi _{-},x\right)+\ldots \right],
\label{Fact.Sec.04.03b}
\end{align}
%%%%%%%%%%%%%%%%%%%%%%%%%%%%%%%%%%%%%%%%%%%%%%%%%%%%%%%%%%%%%%%%%%%%%%%%%%%%%%%
where we choose arbitrary tensors at each order such that $a^{2}_{+}f_{\mu \nu}\left(\phi _{+},x\right)=a^{2}_{-}f_{\mu \nu}\left(\phi _{-},x\right)$ and $a^{2}_{+}q_{\mu \nu}\left(\phi _{+},x\right)=a^{2}_{-}q_{\mu \nu}\left(\phi _{-},x\right)$. Imposing all these conditions we finally obtain
%%%%%%%%%%%%%%%%%%%%%%%%%%%%%%%%%%%%%%%%%%%%%%%%%%%%%%%%%%%%%%%%%%%%%%%%%%%%%%
\begin{align}\label{Fact.Sec.04.04}
h_{\mu \nu}\left(\phi _{-},x\right)=\Omega ^{2}h_{\mu \nu}\left(\phi _{+},x\right);
\qquad
\Omega ^{2}=a^{2}\left(\phi _{-}\right)/a^{2}\left(\phi _{+}\right)=\exp \left[\frac{2\left(\phi _{+}-\phi _{-}\right)}{\ell}\right].
\end{align}
%%%%%%%%%%%%%%%%%%%%%%%%%%%%%%%%%%%%%%%%%%%%%%%%%%%%%%%%%%%%%%%%%%%%%%%%%%%%%%%%%%5
Note that since the branes are not fixed the factor $\Omega$ depends on brane coordinates. Also $\Omega$ depends only on the separation, $\phi _{-}-\phi _{+}$, i.e. on the radion field. Thus we observe that in general for any order in the gradient expansion scheme, we can have the relation (\ref{Fact.Sec.04.04}), where metric on the brane located at $y=\phi _{-}$ is connected to the metric on the brane located at $y=\phi _{+}$ by a conformal factor. Thus the Ricci tensor, the Ricci scalar and the Einstein tensor in the two branes are related through the following relation:
%%%%%%%%%%%%%%%%%%%%%%%%%%%%%%%%%%%%%%%%%%%%%%%%%%%%%%%%%%%%%%%%%%%%%%%%%%%%%%%%%%%%55
\begin{subequations}
\begin{align}
R_{\mu \nu}^{-}&=R_{\mu \nu}^{+}+\frac{1}{\Omega ^{2}}\nabla _{\mu}\nabla _{\nu}\Omega ^{2}
-\frac{3}{2\Omega ^{4}}\nabla _{\mu}\Omega ^{2}\nabla _{\nu}\Omega ^{2}
+\frac{1}{2\Omega ^{2}}h_{\mu \nu}^{+}\nabla _{\alpha}\nabla ^{\alpha}\Omega ^{2}
\label{Fact.Sec.04.05a}
\\
R^{-}&=\frac{1}{\Omega ^{2}}\left[R^{+}+\frac{3}{\Omega ^{2}}\nabla _{\mu}\nabla ^{\mu}\Omega ^{2}
-\frac{3}{2\Omega ^{4}}\nabla _{\mu}\Omega ^{2}\nabla ^{\mu}\Omega ^{2}\right]
\label{Fact.Sec.04.05b}
\\
G^{(-)\mu}_{\nu}&=G^{(+)\mu}_{\nu}+\mathcal{M}^{\mu}_{\nu}
\nonumber
\\
&=G^{(+)\mu}_{\nu}+\frac{1}{\Omega ^{2}}\nabla _{\nu}\nabla ^{\mu}\Omega ^{2}
-\frac{3}{2\Omega ^{4}}\nabla _{\nu}\Omega ^{2}\nabla ^{\mu}\Omega ^{2}
-\delta ^{\mu}_{\nu}\frac{1}{\Omega ^{2}}\nabla _{\alpha}\nabla ^{\alpha}\Omega ^{2}
+\delta ^{\mu}_{\nu}\frac{3}{4\Omega ^{4}}\nabla _{\alpha}\Omega ^{2}\nabla ^{\alpha}\Omega ^{2},
\label{Fact.Sec.04.05c}
\end{align}
\end{subequations} 
%%%%%%%%%%%%%%%%%%%%%%%%%%%%%%%%%%%%%%%%%%%%%%%%%%%%%%%%%%%%%%%%%%%%%%%%%%%%%%%%%%5
where the object $\mathcal{M}^{\mu}_{\nu}$ is defined through \eq{Fact.Sec.04.05c}. We, therefore, have the following Einstein's equation on the two branes:
%%%%%%%%%%%%%%%%%%%%%%%%%%%%%%%%%%%%%%%%%%%%%%%%%%%%%%%%%%
\begin{subequations}
\begin{align}
\frac{\ell}{2}G^{(+)\mu}_{\nu}&=\frac{\kappa ^{2}}{2}T^{(+)\mu}_{\nu}
\label{Fact.Sec.04.06a}
\\
\frac{\ell}{2}G^{(-)\mu}_{\nu}&=\frac{\ell}{2}\left(G^{(+)\mu}_{\nu}+\mathcal{M}^{\mu}_{\nu}\right)
=\frac{\kappa ^{2}}{2}T^{(-)\mu}_{\nu}
\label{Fact.Sec.04.06b}
\end{align}
\end{subequations}
%%%%%%%%%%%%%%%%%%%%%%%%%%%%%%%%%%%%%%%%%%%%%%%%%%%%%%%%%%%%%%%%%%%%%%%%%%%%%%%5
Thus, we observe
%%%%%%%%%%%%%%%%%%%%%%%%%%%%%%%%%%%%%%%%%%%%%%%%%%%%%%%%%%%%%%%%%%%%%%%%%%%%%5
\begin{align}\label{Fact.Sec.04.07}
\frac{\kappa ^{2}}{\ell}\left[\frac{1}{\Psi}T^{(+)\mu}_{\nu}-\frac{1-\Psi}{\Psi}T^{(-)\mu}_{\nu}\right]
=\frac{1}{\Psi}G^{(+)\mu}_{\nu}-\frac{1-\Psi}{\Psi}\left(G^{(+)\mu}_{\nu}+\mathcal{M}^{\mu}_{\nu}\right)
=G^{(+)\mu}_{\nu}-\frac{1-\Psi}{\Psi}\mathcal{M}^{\mu}_{\nu},
\end{align}
%%%%%%%%%%%%%%%%%%%%%%%%%%%%%%%%%%%%%%%%%%%%%%%%%%%%%%%%%%%%%%%%%%%%5
where we have defined $\Psi =1-\Omega ^{2}$. The field equation now leads to the following form:
%%%%%%%%%%%%%%%%%%%%%%%%%%%%%%%%%%%%%%%%%%%%%%%%%%%%%%%%%%%%%%%%%%%%%%%%%5
\begin{align}\label{Fact.Sec.04.08}
G^{(+)\mu}_{\nu}&=\frac{\kappa ^{2}}{\ell}\left[\frac{1}{\Psi}T^{(+)\mu}_{\nu}
-\frac{1-\Psi}{\Psi}T^{(-)\mu}_{\nu}\right]+\frac{1-\Psi}{\Psi}\mathcal{M}^{\mu}_{\nu}
\nonumber
\\
&=\frac{\kappa ^{2}}{\ell}\left[\frac{1}{\Psi}T^{(+)\mu}_{\nu}
-\frac{1-\Psi}{\Psi}T^{(-)\mu}_{\nu}\right]
\nonumber
\\
&-\frac{1}{\Psi}\left[\left(\nabla ^{\mu}\nabla _{\nu}\Psi -\delta ^{\mu}_{\nu}\nabla ^{\alpha}\nabla _{\alpha}\Psi \right)+\frac{\omega \left(\Psi \right)}{\Psi}\left(\nabla ^{\mu}\nabla _{\nu}\Psi
-\frac{1}{2}\delta ^{\mu}_{\nu}\nabla _{\alpha}\Psi \nabla ^{\alpha}\Psi \right)\right],
\end{align}
%%%%%%%%%%%%%%%%%%%%%%%%%%%%%%%%%%%%%%%%%%%%%%%%%%%%%%%%%%%%%%%%%%%%%%%%%5
where we have introduced a new function, $\omega (\Psi)=3\Psi/2(1-\Psi)$. Again eliminating $G^{(+)\mu}_{\nu}$ from \eqs{Fact.Sec.04.06a} and (\ref{Fact.Sec.04.06b}) with the contraction of the indices, we arrive at
%%%%%%%%%%%%%%%%%%%%%%%%%%%%%%%%%%%%%%%%%%%%%%%%%%%%%%%%%%%%%%%%%%%%%%%%%%%%%%%
\begin{equation}\label{Fact.Sec.04.09}
\square \Psi +\frac{2\omega +3}{3}\nabla _{\alpha}\Psi \nabla ^{\alpha}\Psi 
=\frac{\kappa ^{2}}{\ell}\frac{1}{2\omega +3}\left(T^{(-)\mu}_{\mu}-T^{(+)\mu}_{\mu}\right)
\end{equation}
%%%%%%%%%%%%%%%%%%%%%%%%%%%%%%%%%%%%%%%%%%%%%%%%%%%%%%%%%%%%%%%%%%%%%%%%%%%%%%%%%%5
Note that the field equation for gravity given in \eq{Fact.Sec.04.08} and the field equation for $\Psi$ provided by \eq{Fact.Sec.04.09} hold for any order in the gradient expansion scheme. Thus, the field equation for $\Psi$, equivalently, for the radion field, is determined by the trace of the stress energy tensor at both branes. The remarkable thing about these field equations are that they hold for all orders in the perturbation scheme and is equivalent to Brans-Dicke field equations for gravity. 

In order to make the circle complete let us write down the effective equation entirely in terms of $\Omega$. For that we note the following identities,
%%%%%%%%%%%%%%%%%%%%%%%%%%%%%%%%%%%%%%%%%%%%%%%%%%%%%%%%%%%%%%%%%%%%%%%%%%%%%%%%%%%%
\begin{align}
\partial _{a}\Omega ^{2}&=\Omega ^{2}\frac{2}{\ell}\left(\partial _{a}\phi _{+}-\partial _{a}\phi _{-}\right)
\label{Fact.Sec.04.10a}
\\
\partial _{a}\Omega ^{2}\partial ^{a}\Omega ^{2}&=\frac{4\Omega ^{4}}{\ell ^{2}}\left[
\left(\partial _{a}\phi _{+}\right)^{2}+\left(\partial _{a}\phi _{-}\right)^{2}
-2\partial _{a}\phi _{-}\partial ^{a}\phi _{+}\right]
\label{Fact.Sec.04.10b}
\\
\phi _{+}-\phi _{-}&=\frac{\ell}{2}\ln \Omega ^{2}.
\label{Fact.Sec.04.10c}
\end{align} 
%%%%%%%%%%%%%%%%%%%%%%%%%%%%%%%%%%%%%%%%%%%%%%%%%%%%%555555%%%%%%%%%%%%%%%%%%%%55
However, in order to get a clear picture we set $\phi _{+}=\textrm{constant}$, such that $a^{2}_{+}=1$. Thus, up to second order, the effective equation turns out to have the following form from \eq{Fact.Sec.03.13}:
%%%%%%%%%%%%%%%%%%%%%%%%%%%%%%%%%%%%%%%%%%%%%%%%%%%%%%%%%%%%%%%%%%%%%%%%%%%%%%%%%5
\begin{align}\label{Fact.Sec.04.11}
S_{\rm{tot}}&=\frac{\ell}{2\kappa ^{2}}\int d^{4}x\sqrt{-g}\Big[\left(1-\Omega ^{2}\right)R
-\frac{3}{2\Omega ^{2}}\partial _{\mu}\Omega ^{2}\partial ^{\mu}\Omega ^{2}
-\frac{3\ell ^{5}}{64}\ln \Omega ^{2}
\left(R^{\alpha}_{\beta}R^{\beta}_{\alpha}-\frac{1}{3}R^{2}\right)\Big]
\nonumber
\\
&=\frac{\ell}{2\kappa ^{2}}\int d^{4}x\sqrt{-g}\Big[\Psi R
-\frac{\omega \left(\Psi \right)}{\Psi}\partial _{\mu}\Psi\partial ^{\mu}\Psi
-\frac{3\ell ^{5}}{64}\ln \Omega ^{2}
\left(R^{\alpha}_{\beta}R^{\beta}_{\alpha}-\frac{1}{3}R^{2}\right)\Big].
\end{align}
%%%%%%%%%%%%%%%%%%%%%%%%%%%%%%%%%%%%%%%%%%%%%%%%%%%%%%%%%%%%%%%%%%%%%%%%%%%%%%%%%%%%%%%%%%%
which resembles the action for Brans-Dicke theory of gravity. Thus, even at the level of the effective action the bulk-brane system is equivalent to Brans-Dicke or scalar-tensor theories of gravity.

We should stress that in \cite{Kanno2003} the equivalence with Brans-Dicke theory was shown for first order in the gradient expansion scheme. In this work we have shown explicitly that the effective action with second-order corrections included resembles the Brans-Dicke theory of gravity. However, our argument uses arbitrary tensors at each order and, thus, holds for any order in the gradient expansion scheme. Therefore, the resemblance of brane world model with the Brans-Dicke theory of gravity holds at \emph{all} orders in the gradient expansion scheme. 
%%%%%%%%%%%%%%%%%%%%%%%%%%%%%%%%%%%%%%%%%%%%%%%%%%%%%%%%%%%%%%%%%%%%%%%%%%%%%%%%%%%%%%%%%%%%%%%%%%%%%5
%%%%%%%%%%%%%%%%%%%%%%%%%%%%%%%%%%%%%%%%%%%%%%%%%%%%%%%%%%%%%%%%%%%%%%%%%%%%%%%%%%%%%%%%%%%%%%%%%%%%5
%%%%%%%%%%%%%%%%%%%%%%%%%%%%%%%%%%%%%%%%%%%%%%%%%%%%%%%%%%%%%%%%%%%%%%%%%%%%%%%%%%%%%%%%%%%%%%%%%%%%%%5
\section{Discussion}

In this work, our main aim was to address two important aspects related to brane world models. First, the issue of factorizability of the metric ansatz, and second, the equivalence of Brans-Dicke theory with this brane world model. Previous steps in these directions were taken in \cite{Kanno2002a,Kanno2002b,Kanno2003,Kanno2004,Kanno2005}. Our work, however, generalizes their results and relaxes most their assumptions. The key results in our analysis, which differ significantly from those presented in earlier attempts, can be summarized as follows:
%%%%%%%%%%%%%%%%%%%%%%%%%%%%%%%%%%%%%%%%%%%%%%%%%%%%%%%%
%%%%%%%%%%%%%%%%%%%%%%%%%%%%%%%%%%%%%%%%%%%%%%%%%%%%%%%%
\begin{itemize}
%%%%%%%%%%%%%%%%%%%%%%%%%%%%%%%%%%%%%%%%%%%%%%%%%%%%%%%%
\item In \cite{Kanno2002b} the second-order corrections in the gradient expansion scheme were calculated however with two assumptions: (a) branes are located at fixed positions and (b) higher order terms of the arbitrary tensors can be neglected. In this work we have generalized the previous result to second order in the gradient expansion scheme, by deriving the second order correction to the bulk metric by relaxing both these assumptions. We have taken the brane position to be variable and have included all the corrections originating from the arbitrary tensors in our calculation.
%%%%%%%%%%%%%%%%%%%%%%%%%%%%%%%%%%%%%%%%%%%%%%%%%%%%%%%%
\item From our work it turns out that the effective action gets factorized into the extra dimension or radion part and brane part even when the second order corrections are included (this generalizes previous results derived only up to first order \cite{Kanno2005}). Also by generalizing our result we can argue that factorizability is a valid assumption upto \emph{all} order in this perturbative expansion scheme.
%%%%%%%%%%%%%%%%%%%%%%%%%%%%%%%%%%%%%%%%%%%%%%%%%%%%%%%
\item Non-local factors originating from the bulk field equations can be used to express the gravitational field equation on the brane in terms of the radion field and bulk metric. This has been done earlier in \cite{Kanno2003} however with only the first order corrections to the effective equation. In this work we have incorporated the second order corrections and have devised a generic way which can easily be extended to any orders in the gradient expansion scheme.
%%%%%%%%%%%%%%%%%%%%%%%%%%%%%%%%%%%%%%%%%%%%%%%%%%%%%%%
\item Through this work we can conclude that the two brane system is equivalent to Brans-Dicke theory as far as the effective description is considered and this is true for \emph{all} orders in the perturbative gradient expansion valid at low energies.
%%%%%%%%%%%%%%%%%%%%%%%%%%%%%%%%%%%%%%%%%%%%%%%%%%%%%%%
\end{itemize}
%%%%%%%%%%%%%%%%%%%%%%%%%%%%%%%%%%%%%%%%%%%%%%%%%%%%%%%%
%%%%%%%%%%%%%%%%%%%%%%%%%%%%%%%%%%%%%%%%%%%%%%%%%%%%%%%%
Hence our work shows that metric factorizability is a valid assumption in \emph{all} orders of the perturbation theory with the ratio of four dimensional curvature to the five dimensional one as a perturbative parameter. Secondly, we were able to show that brane world model is equivalent to Brans-Dicke theory of gravity. This is also true in \emph{all} orders of the perturbative expansion. Thus we can conclude that metric factorizability and equivalence of brane word models with Brans-Dicke theory holds in low energy (i.e., when brane to bulk curvature ratio is small) to \emph{all} orders in gradient expansion scheme which generalizes all the previous results in this direction. %%%%%%%%%%%%%%%%%%%%%%%%%%%%%%%%%%%%%%%%%%%%%%%%%%%%%%%%%%%%%%%%%%%%%%%%%%%%%%%%%%%%%%%%%%%%%%%%%%%%%
%%%%%%%%%%%%%%%%%%%%%%%%%%%%%%%%%%%%%%%%%%%%%%%%%%%%%%%%%%%%%%%%%%%%%%%%%%%%%%%%%%%%%%%%%%%%%%%%%%%%%%
%%%%%%%%%%%%%%%%%%%%%%%%%%%%%%%%%%%%%%%%%%%%%%%%%%%%%%%%%%%%%%%%%%%%%%%%%%%%%%%%%%%%%%%%%%%%%%%%%%%%%%
\section*{Acknowledgement}

S.C. thanks IACS, India for warm hospitality; a part of this work was completed there during a visit. He also thanks CSIR, Government of India, for providing a SPM fellowship.

%%%%%%%%%%%%%%%%%%%%%%%%%%%%%%%%%%%%%%%%%%%%%%%%%%%%%%%%%%%%%%%%%%%%%%%%%%%%%%%%%%%%%%%%%%%%%%%%%%%%%%%
%%%%%%%%%%%%%%%%%%%%%%%%%%%%%%%%%%%%%%%%%%%%%%%%%%%%%%%%%%%%%%%%%%%%%%%%%%%%%%%%%%%%%%%%%%%%%%%%%%%%%%%
%%%%%%%%%%%%%%%%%%%%%%%%%%%%%%%%%%%%%%%%%%%%%%%%%%%%%%%%%%%%%%%%%%%%%%%%%%%%%%%%%%%%%%%%%%%%%%%%%%%%%%%

%%%%%%%%%%%%%%%%%%%%%%%%%%%%%%%%%%%%%%%%%%%%%%%%%%%%%%%%%%%%%%%%%%%%%%%%%%%%%%%%%%%%%%%%%%%%%%%%%%%%%%5
%%%%%%%%%%%%%%%%%%%%%%%%%%%%%%%%%%%%%%%%%%%%%%%%%%%%%%%%%%%%%%%%%%%%%%%%%%%%%%%%%%%%%%%%%%%%%%%%%%%%%%
%%%%%%%%%%%%%%%%%%%%%%%%%%%%%%%%%%%%%%%%%%%%%%%%%%%%%%%%%%%%%%%%%%%%%%%%%%%%%%%%%%%%%%%%%%%%%%%%%%%%%%%5
\end{document}